\documentclass[pra,twocolumn, floatfix, footinbib,superscriptaddress, longbibliography]{revtex4-1}
\usepackage[utf8x]{inputenc}
\usepackage[english]{babel}
\usepackage[english]{babel}
\usepackage[colorlinks=true,bookmarks=false,citecolor=blue,urlcolor=blue]{hyperref}

\usepackage{amsmath,amssymb,esint}
\usepackage{graphicx}
\usepackage{verbatim}
\usepackage{color}
\usepackage[english]{babel}
\usepackage{bm}
\usepackage{natbib}
\usepackage[normalem]{ulem}
\usepackage{epstopdf}
\usepackage{multirow}
\usepackage[dvipsnames]{xcolor}
\usepackage{ragged2e}

\renewcommand {\phi}{\varphi}

\usepackage{amsmath}
\usepackage{graphicx}
\usepackage[colorinlistoftodos]{todonotes}
\newcommand{\veps}{{\varepsilon}}

\newcommand{\bee}{\begin{equation}}
\newcommand{\ene}{\end{equation}}

\newcommand{\bea}{\begin{eqnarray}}
\newcommand{\ena}{\end{eqnarray}}

\graphicspath{{pics/}}

\let\vec=\mathbf

\begin{document}
\title{Multipolar origin of bound states in the continuum}

\author{Zarina Sadrieva}
\email{z.sadrieva@metalab.ifmo.ru}
\affiliation{ITMO University, St. Petersburg 197101, Russia}

\author{Kristina Frizyuk}%\normalfont\textsuperscript{a,}
\email{k.frizyuk@metalab.ifmo.ru}
\affiliation{ITMO University, St. Petersburg 197101, Russia}

\author{Mihail Petrov}
\affiliation{ITMO University, St. Petersburg 197101, Russia}

\author{Yuri Kivshar}
\affiliation{ITMO University, St. Petersburg 197101, Russia}
\affiliation{Nonlinear Physics Center, Australian National University, Canberra ACT 2601, Australia}

\author{Andrey Bogdanov}
\affiliation{ITMO University, St. Petersburg 197101, Russia}
%Department of Nanophotonics and Metamaterials, ITMO University, Saint Petersburg 197101, Russia
%Nonlinear Physics Center, Australian National University, Canberra, Australian Capital Territory 2601, Australia

\begin{abstract}
Metasurfaces based on resonant subwavelength photonic structures enable novel ways of wavefront control and light focusing, underpinning a new generation of flat-optics devices. Recently emerged all-dielectric metasurfaces exhibit high-quality resonances underpinned by the physics of bound states in the continuum that drives many interesting concepts in photonics. Here we suggest a novel approach to explain the physics of bound photonic states embedded into the radiation continuum.  We study dielectric metasurfaces composed of planar periodic arrays of Mie-resonant  nanoparticles ("meta-atoms") which support both symmetry protected and accidental bound states in the continuum, and employ the multipole decomposition approach to reveal the physical mechanism of the formation of such nonradiating states in terms of multipolar modes generated by isolated meta-atoms. Based on the symmetry of the vector spherical harmonics, we identify the conditions for the existence of bound  states in the continuum originating from the symmetries of both the lattice and the unit cell. Using this formalism we predict that metasurfaces with strongly suppressed spatial dispersion can support the bound states in the continuum with the wavevectors forming a line in the reciprocal space. Our results provide a new way for designing high-quality resonant photonic systems based on the physics of bound states in the continuum.
\end{abstract}

\keywords{nanophotonics, multipolar decomposition, metasurface, photonic crystals, bound states in the continuum}

\pacs{}

\maketitle

\section{Introduction}

The quest for compact photonic systems with high quality factor ($Q$ factor) modes led to the rapid development of  optical bound states in the continuum (BICs). BICs are non-radiating states, characterized by the resonant frequencies embedded to the continuum spectrum of radiating modes of the surrounding space~\cite{Hsu2016, KOSHELEV62018}. The BICs first appeared as a mathematical curiosity in quantum mechanics~\cite{Neumann1929}. The discovery of BICs in optics immediately attracted  broad attention (see, e.g., Refs.~\cite{Paddon2000,Bulgakov2008, Marinica2008}) due to high potential in applications in communications~\cite{Dreisow2009,Gentry2014}, lasing~\cite{Kuznetsov2018,Lepetit2017,penzo2017,rybin2017}, filtering~\cite{foley2015}, and sensing~\cite{Foley2014,romano2018Materials,romano2018OSA}. Recent achievements in the field of BIC are discussed in Refs.~\cite{NonlKoshelev,Gao2016,Bulgakov2014,Shalaev,qfactorarxiv,chen2019,abujetas2019,kartashov2018,romano2018,bulgakov2017}. %Recently, the quasi-BIC in an isolated subwavelength cylinder has been demonstrated~\cite{RybinPRL}.

Decoupling of the resonant mode from the radiative spectrum, which is the basic idea behind the BIC, can be interpreted in several equivalent ways. Within the coupled-mode theory, it corresponds to nullifying  the coupling coefficient between the resonant mode and all radiation channels of the surrounding space~\cite{WeiHsu2013}. Alternatively,  the appearance of BICs is explained as vanishing of Fourier coefficients corresponding to open diffraction channels due to the symmetry of the photonic structure. At the particular high-symmetry points of the reciprocal space, for example, at the $\Gamma$ point, the continuous spectrum is divided into the modes of different symmetry classes with respect to the reflectional and rotational symmetry of the photonic system. The bound states of one symmetry class can be found embedded in the continuum of another symmetry class, and their coupling is forbidden as long as the symmetry is preserved. Such kind of BIC is called {\it symmetry-protected}, and they  also allow interpretation in terms  of topological charges defined by the number of times the polarization vectors winds around the BICs presented as vortex centers in the polarization field~\cite{Zhen2014}. In contrast to the symmetry-protected BIC, the so-called {\it accidental BICs}~\cite{Hsu2013,Zhen2014,Hsu2016, groupTuz, PhysRevB.94.245148} can be observed out of the $\Gamma$-point due to an accidental nulling of the Fourier (coupling) coefficients via fine tuning of parameters of the photonic system. Such a mechanism is also known as Friedrich-Wintgen scenario~\cite{Friedrich}.
%They do not posses robustness to a small variation of the photonic structure parameters but still can be  observed experimentally.

%========================Figure============================================================
 \begin{figure}
	\includegraphics[width=1\linewidth]{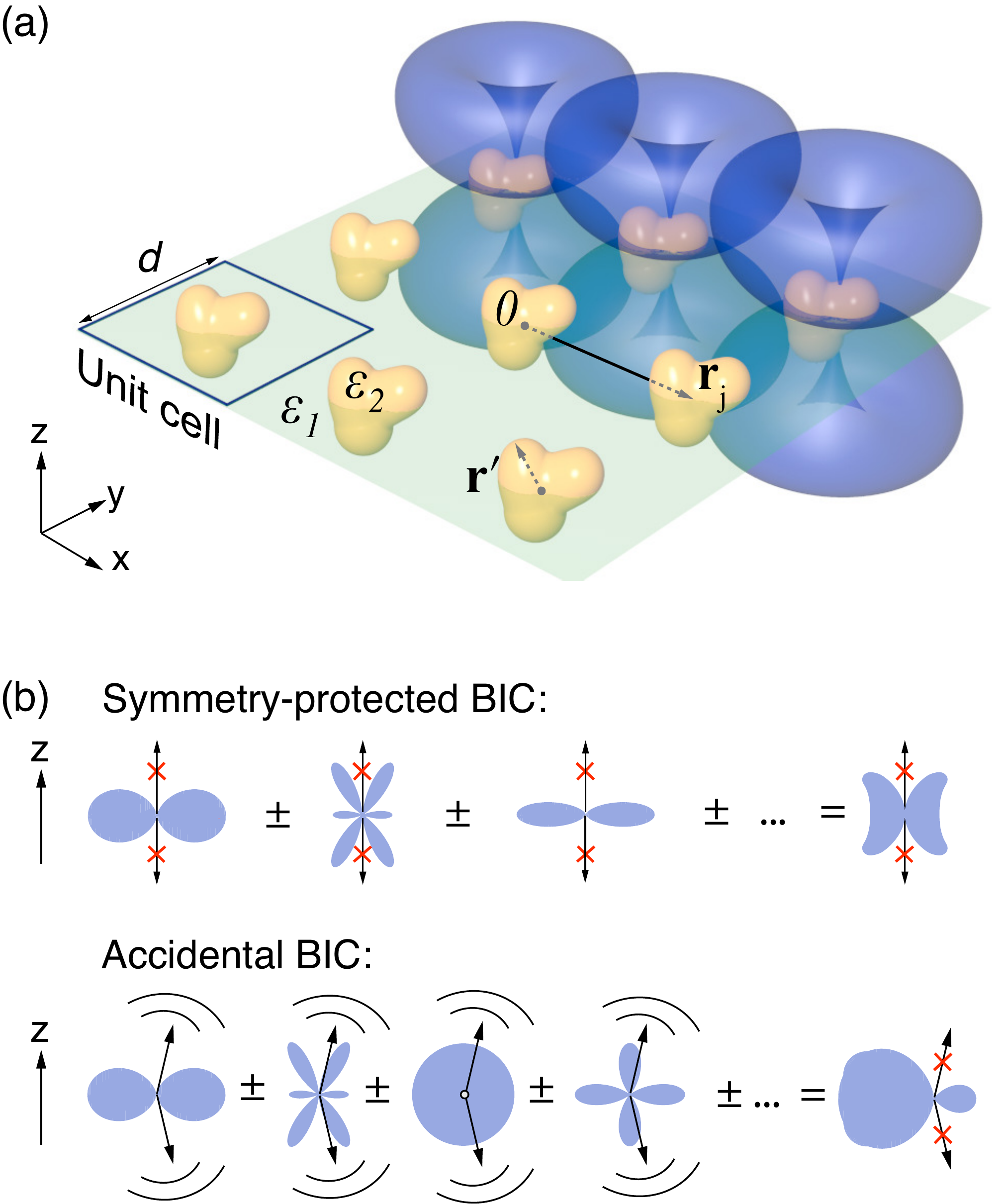}
	 \caption{(a) Metasurface with square unit cell composed of dielectric meta-atoms. (b) The upper and lower panels show the formation mechanism of symmetry protected (at-$\Gamma$) and accidental (off-$\Gamma$) BIC.} \label{figmeta}% The period of the array $d=600$~nm, radius of the nanosphere is $a=100$~nm and dielectric permittivity of the nanosphere material is $\varepsilon_2=12$, $\varepsilon_1=1$.}
\end{figure}
%========================Figure ends============================================================

Despite the number of existing approaches to understand the nature of BICs, there is still a room for further development of the theory. During the previous few years the electromagnetic multipole theory~\cite{Jackson1965} has been extensively developed as a natural tool of nanophotonics  dealing with the lowest (fundamental) resonances of the system. The main advantage of the multipole decomposition method (MDM) is that it provides a representation of an arbitrary field distribution as a superposition of the fields created by a set of multipoles~\cite{Grahn2012, Kruk2017}. Namely, the multipole expansion has been widely used to determine the polarization and directivity patterns of the scattered field of single particles and their clusters both plasmonic and dielectric~\cite{PRBEvlykhin,Grahn2012} for a variety of applications such as polarization control device~\cite{Kruk2017}, dielectric nanoantenna~\cite{Kuznetsov2016}, light demultiplexing~\cite{Panmai2018}, and others. A number of novel optical phenomena have been explained within the MDM such as  anapole effect~\cite{KsuPRB, anapole2018}, optomechanical phenomena~\cite{Hsu2015, Poshakinskiy}, and Kerker effect~\cite{KsuarXiv,Liu2018}.

In this work, we extend the MDM approach for explaining both symmetry protected and accidental BICs. We provide a theory of BICs origin in terms of MDM for a general case of any periodic structure  and develop an analytical method, which determines the contribution of the vector spherical harmonics to the far field (Section~\ref{formulae}). Working in the vector spherical harmonics (VSH) basis, we take the advantage of the internal symmetry and provide the group-symmetry approach to identifying the BIC formation in terms of the unit cell and lattice symmetries (Section~\ref{groupt}).  We implement field multipole expansion of the eigenmodes of a periodic two-dimensional (2D) photonic structure supporting BIC. We illustrate the developed technique by considering a 2D square array of spheres and extending it to the case of a photonic crystal slab with a 2D array of cylindrical holes (Section~\ref{comss}).  The developed approach can be easily extended even further to periodic structures with other types of the unit cell and other lattice symmetries. The proposed method both provides a deeper understanding of the photonic BIC physics and gives a tool for an effective designing of high-$Q$ resonant photonic systems.

\section{Multipolar approach}  \label{formulae}

In this section we consider the modes of {a two-dimensional periodic array of dielectric nanoparticles {with arbitrary shape} (see Fig.~\ref{figmeta}), }% of dielectric nanospheres.
and obtain an expression connecting the multipolar content of the field {inside and outside the nanoparticles. }%(\red{in general case, outside any of the spheres circumscribed of a unit cell})\commentMisha{its not very clear what does  "outside" mean..}.
We denote the  VSHs as $\vec M_{p_rmn}(k,\vec r)$ (magnetic) and  $\vec N_{p_rmn}(k,\vec r)$ (electric) relying on the  definition presented  in Ref.~\onlinecite{Bohren}. We introduce an additional notation $\vec W_{p_ip_rmn}$ for the both types of VSHs, where inversion parity index $p_i=(-1)^{n+1}$ for  $\vec M_{p_rmn}(k,\vec r)$, and $p_i=(-1)^n$ for $\vec N_{p_rmn}(k,\vec r)$.
Index $n$ is the multipole order,  $m$ varies  from 0 to  $n$, and $p_r=1$ if $\vec W_{p_ip_rmn}$ is even under reflection from $y=0$ plane ($\varphi \rightarrow -\varphi$ in the spherical system), and  $p_r=-1$ if it is odd \cite{frizyuk_arxiv}. %\commentMisha{Got lost with the notation, which index numerates the magnetic/electric part? What does it mean "$p_i=(-1)^{n+1}$ for magnetic harmonics"?}

In homogeneous medium with permittivity $\varepsilon$, any solution of the Helmholtz equation with wavevector  $k=\sqrt{\varepsilon} \frac{\omega}{c}$  can be expanded in terms of electric and magnetic spherical harmonics \cite{Stratton}. The field inside the medium of $j$-th cell of the array $\vec E^{in}$ can be written in terms of multipolar decomposition:
\begin{equation} \label{ein}
 \vec E^{in}(\vec r) =
 \sum_{n=1}^\infty \sum_{^{m=0}_{p_i,p_r}}^n E_0[D_{p_ip_rmn} \vec W^{(1)}_{p_ip_rmn} (k_2, \vec r') e^{i(\vec k_b \cdot \vec r_j) }],
 \end{equation}
where $k_2 = \sqrt{\varepsilon_2} \frac{\omega}{c} $ is the wavevector in the material, the superscript $(1)$ stands for spherical Bessel functions in the radial part of the VSH, $\vec k_b$ is {the} Bloch vector, $\vec r'= \vec r-\vec r_j$, and $\vec r_j$~is the position of a single sphere (see Fig.~\ref{figmeta}).  { $D_{p_ip_rmn}$ are the coefficients of the multipolar decomposition, which will be discussed in Section~\ref{groupt}.}
% \blue {The coefficients of multipolar decomposition $D_{p_ip_rmn}$ turn out to be zero for the modes of particular symmetry, see Section~\ref{groupt} for details.}
In similar manner, the field outside the array is expressed as follows (see Appendix~\ref{Weyl}):
%n our considerations, we use vector spherical harmonics(VSH) $\vec M_{p_rmn}(k,\vec r)$ (magnetic) and  $\vec N_{p_rmn}(k,\vec r)$(electric), defined in \cite{Bohren}. Any solution of the Helmholtz equation with wavevector  $k=\sqrt{\varepsilon} \frac{\omega}{c}$  can be expanded in terms of these functions \cite{Stratton}.
%We introduce the notation $\vec W_{p_ip_rmn}$, for both types of VSH, where $p_i=(-1)^{n+1}$ for magnetic harmonics, and $p_i=(-1)^n$ for electric ones.
%Index $n$ is the multipole order,  $m$ is  from 0 to  $n$, and $p_r=1$ if $\vec W_{p_ip_rmn}$ is even under reflection from $y=0$ plane ($\varphi \rightarrow -\varphi$ in the spherical system), and  $p_r=-1$ if it is odd \cite{frizyuk_arxiv}.

%The field inside the $j$-th sphere of the array $\vec E^{in}$ can be written in terms of multipolar decomposition:
%\begin{equation} \label{ein}
% \vec E^{in}(\vec r) =
% \sum_{n=1}^\infty \sum_{^{m=0}_{p_i,p_r}}^n E_0[D_{p_ip_rmn} \vec W^{(1)}_{p_ip_rmn} (k_2, \vec r') e^{i(\vec k_b \cdot \vec r_j) }],
% \end{equation}
%where $k_2 = \sqrt{\varepsilon_2} \frac{\omega}{c} $ is wavevector in the material, superscript $(1)$ stands for spherical bessel functions in the radial part of the VSH, $\vec k_b$ is Bloch vector, $\vec r'= \vec r-\vec r_j$, and $\vec r_j$~is the position of the single sphere (see figure \ref{figmeta}).
 \begin{alignat}{2}
\label{eq:out202}
\vec E (\vec r)=E_0\sum_{\vec K, p_i, p_r,m,n}  \tilde D_{p_ip_rm n} \frac{V_b i^{-n}}{2\pi k_1}
 \times  \nonumber \\
 \times\iint  \limits_{-\infty}^{\infty} dk_{\parallel}  \delta({\vec k_{b} - \vec k_{\parallel}-\vec K})\frac{e^{ i{\vec k \vec r}}}{k_z}\left[\vec Y_{p_ip_rmn}\left(\frac{\vec k}{|\vec k|}\right)\right].
\end{alignat}
Here $\vec K$ is the reciprocal lattice vector, $k_z~=\pm\sqrt{k_1^2-k_{\parallel}^2}$, $ |\vec k|=k_1= \sqrt{\varepsilon_1} \frac{\omega}{c}$, {$V_b$ is the volume of the first Brillouin zone}. The spherical vector functions $\vec Y(\vec n)$ depend on the spherical coordinates of a unity  vector $\vec n$ and they are given in the Appendix~\ref{app:definitions}. Their symmetry coincides with the symmetry of $\vec W$-function  with the identical indices. The relation between $D$ and $\tilde D$ is discussed in  Appendix~\ref{appcoef}, and  for the case of an array of spherical particles, {coefficients $\tilde D$ can be }derived analytically. %They have similar angular part as the asymptotics of $\vec W$ with similar indexes.
%========================Figure============================================================
% \begin{figure}
%	\includegraphics[width=0.99\linewidth]{illustration5-eps-converted-to.pdf}
%	%\includegraphics[width=0.75\linewidth]{TM3_Q-eps-converted-to.pdf}
%	 \caption{The schematic radiation pattern of a single unit cell corresponding to  $\Gamma$-BIC and off-$\Gamma$-BIC (accidental) illustrated in terms of multipole decomposition.} \label{fig:ill}
%\end{figure}
%========================Figure ends============================================================

The summation   in Eq.~\eqref{eq:out202}  over the reciprocal vectors corresponding to open diffraction channels, i.e. the terms with real $k_z$, provides the contribution into the far field.  For the frequencies below the diffraction limit, only the zero-order term with $ \vec K=0$ gives non-zero contribution to the far field:
 \begin{alignat}{2}
\vec E (\vec r)=\frac{E_0 V_b }{{2\pi k_1 k_{1z}}} {e^{ i{\vec k_1 \vec r}}}\sum_{^{p_i, p_r,}_{ m, n}}  i^{-n} \tilde D_{p_ip_rmn} \left[\vec Y_{p_ip_rmn}\left(\frac{\vec k_1}{k_1}\right)\right], \label{eq:farfield}
 \end{alignat}
where $\vec k_{1||}=\vec k_b,$ and $k_{1z}=\pm\sqrt{k_1^2-k_{b}^2}$. According to Eq.~\eqref{eq:farfield}, the contribution of the  multipole with numbers $p_i, p_r,m,n$ into the far field in the direction defined by the wave vector { $\vec k_1$} is proportional to the multipole expansion coefficient $\tilde D$ and the value of spherical vector function $\vec Y$ in the given direction. Equation~\eqref{eq:farfield} provides the  correspondence  between the radiation pattern of a single unit cell and the far-field properties of the whole infinite array allowing for
interpreting  the BIC  in terms of MDM. In strong contrast to a single nanoparticle, where each multipole contributes to the far field, in case of a subdiffractive array there might be direction, where non of the multipole gives any contribution, or alternatively the non-zero contribution of different terms may eventually sum up to zero. The formulated alternative gives a sharp distinction between the symmetry protected and accidental BIC.

{\it At-$\Gamma$-point BIC.} The $\Gamma$-point BIC corresponds to the absence of the far-field radiation in the direction along  the $z$-axis. Due to the structure of VSH, it appears that a number of multipoles do not radiate in the vertical direction along the $z$-axis. If the field inside a single unit cell consists only of such multipoles, there will be no total radiation in $z$-direction. This simple fact is illustrated in the upper panel in Fig.~\ref{figmeta}(b).  {Noticing that only $\vec Y_{p_ip_r1n}$  functions with m=1 are non-zero in parallel to the $z-$axis direction, we can conclude that at the $\Gamma$-point in the subdiffractive array all the modes which do not contain the harmonics with $m=1$ are symmetry-protected BICs. This fundamental conclusion lies in the basis of recent experimental demonstration~\cite{Kuznetsov2018} of lasing with BIC in a 2D subdiffractive array of nanoparticles. The particular operational mode consisted of vertical dipoles oriented along the $z$-axis, thus, not contributing into the only open channel. There exist an approach~\cite{Zhen2014,Hsu2016} that the eigenmodes at the $\Gamma$-point can radiate in the normal direction $z$ if their fields are odd under $C^z_2$ rotations, and do not have any other rotational symmetry of $C^z_n$-type. %\sout{For a photonic system with square lattice, odd modes with  respect $C_2^z$  symmetry radiates in the normal direction $z$}.
 In terms of multipole moments, this follows from the fact that at the $\Gamma$-point any radiative mode should contain multipoles with $m=1$. On the other hand, in virtue of the symmetry, the even modes have zero radiation losses, i.e. infinite radiation quality factor, which are known as symmetry-protected BICs.%One can see that BICs modes TE$_1$ , TE$_4$, TM$_3$ and TM$_4$ are characterized by even field distribution. Away from the $\Gamma$-point, modes are no longer protected by symmetry, and BICs happen at an accidental $k$ point, for instance, {for the} TM$_3$ mode, see Fig.~\ref{band}b.}

{\it ~Off-$\Gamma$ BIC.} Let us now turn to the case of accidental  BIC description.
 In general case, coefficients $D$ are complex numbers. They define the amplitudes and phase delay between the multipoles. However, in accordance with \cite{Hsu2013}, if the structure has time reversal and inversion symmetry, the eigenmodes must satisfy the condition $\vec E(\vec r)=\vec E^* (\vec {-r})$. This fact imposes strict conditions on the multipoles' phases, because some of them are even under inversion and some of them are odd, the coefficient $D_{-1p_rmn}$ before the odd ones must be imaginary. %We reflect this in the Table \ref{tableTE} with imaginary unit before some of the harmonics in the valleys.
 %Inversion parity of the multipole  $\vec W_{p_ip_rmn}$ is determined by the index $p_i$. This allows us to rewrite the formula \eqref{eq:farfield}:
 { It follows that every term of t}he sum in Eq.~\eqref{eq:farfield} is purely real, and all multipoles  are in-phase or anti-phase.  {All coefficients depend on k-vector and structure's parameters and in a case of the off-$\Gamma$ BICs this sum turns to zero}. % \textbf{Kristina, we can try to prove it with COMSOL. - We have already proved this kak mogli))}
 {In other words}, for the particular $\vec k_1$ all vector harmonics add up to zero in the direction of $\vec k_1$, analogously to the anti-Kerker effect, because they are already in phase and only amplitudes are modulated while k-vector is changed [Fig. \ref{figmeta}(b), lower panel]. One can intuitively understand accidental BIC (TM-polirized) via a toy dipole model composed of vertical electric and horizontal magnetic dipoles interfering destructively at some k-vector~\cite{doeleman2018}.

The expansion coefficients $ D_{p_ip_rmn}$ depend on shape of the nanoparticles, material parameters and symmetry of the lattice. Obviously, that the lattice symmetry should impose restrictions on $D_{p_ip_rmn}$ and some coefficients should vanish due to the symmetry. To explore these selection rules, we employ theoretical-group approach.

% ********************* Section2******************************
\section{Symmetry approach} \label{groupt}

 \begin{figure}
	\includegraphics[width=0.95\linewidth]{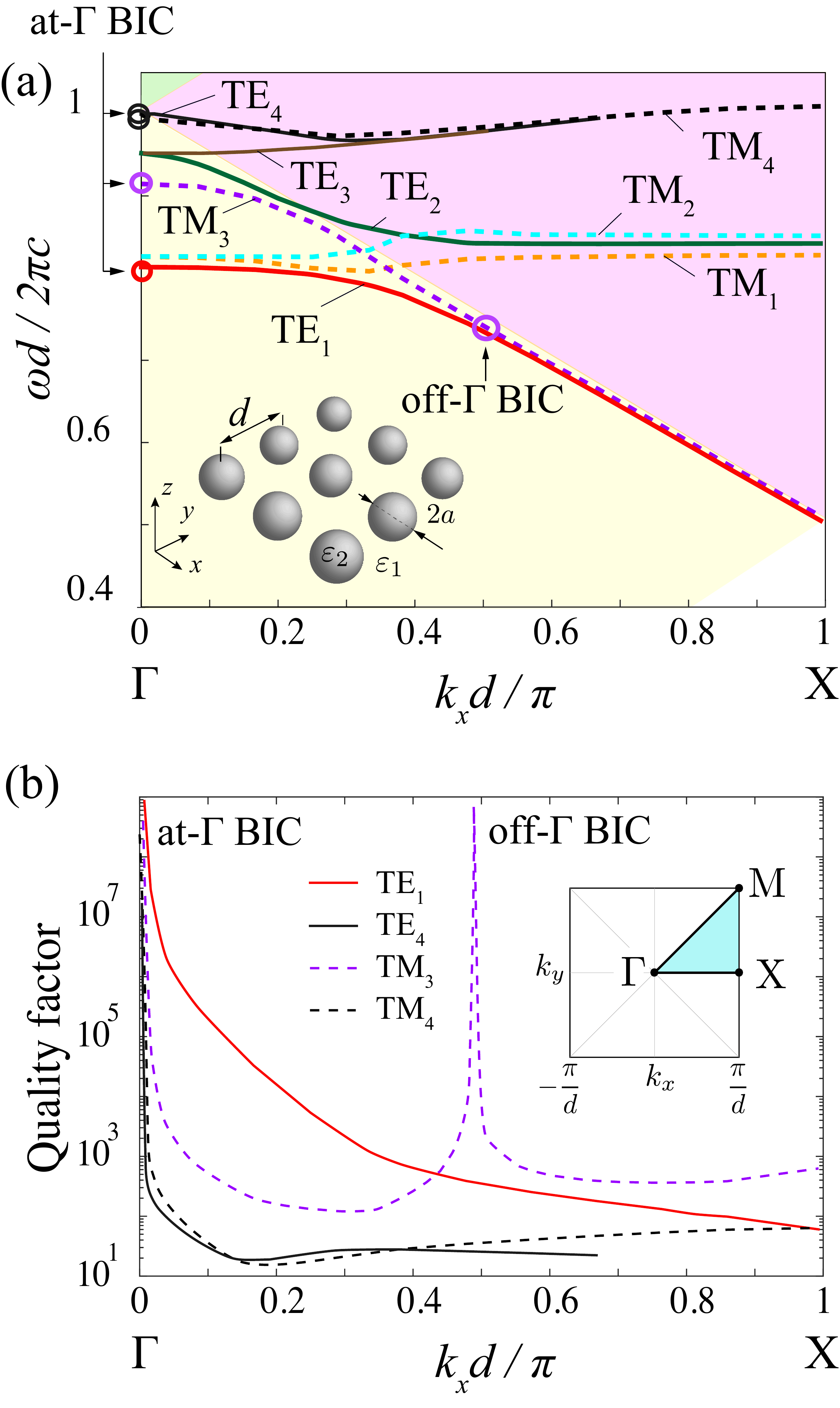}
	 \caption{(a) Band diagram for the square periodic array of dielectric spheres at the $\overline{\Gamma X}$-valley. Period $d$ = 600 nm, nanosphere radius $a$=100nm, dielectric permittivity of the spheres is $\varepsilon_2$=12, medium permittivity is $\varepsilon_1$=1. Modes TE$_1$, TE$_4$, TM$_3$, TM$_4$, are symmetry-protected BICs at the $\Gamma$-point. (b) Q-factor of the TM$_3$ mode in the  $\overline{\Gamma X}$-valley. At the particular wave vector, Q-factor tends to infinity and the accidental BIC appears. } \label{band}
\end{figure}

%========================Table============================================================

 \begin{table}
 	 \caption{Multipolar content and irreducible representations  of TE-modes at the $\Gamma$-point and at $\overline{\Gamma X}$ and $\overline\Gamma M$ valleys. Spherical harmonics for $B_2$ and $A_1$ representations at $\overline{\Gamma X}$ are provided at the right column of the table. The multipoles for the same representations at $\overline{\Gamma M}$ can be obtained by rotation by $\pi/4$.  Imaginary unit means that coefficients $D$ before these harmonics must be imaginary. The $z$-component of the magnetic field $H_z$ is shown in the inserts. }
	\includegraphics[width=0.97\linewidth]{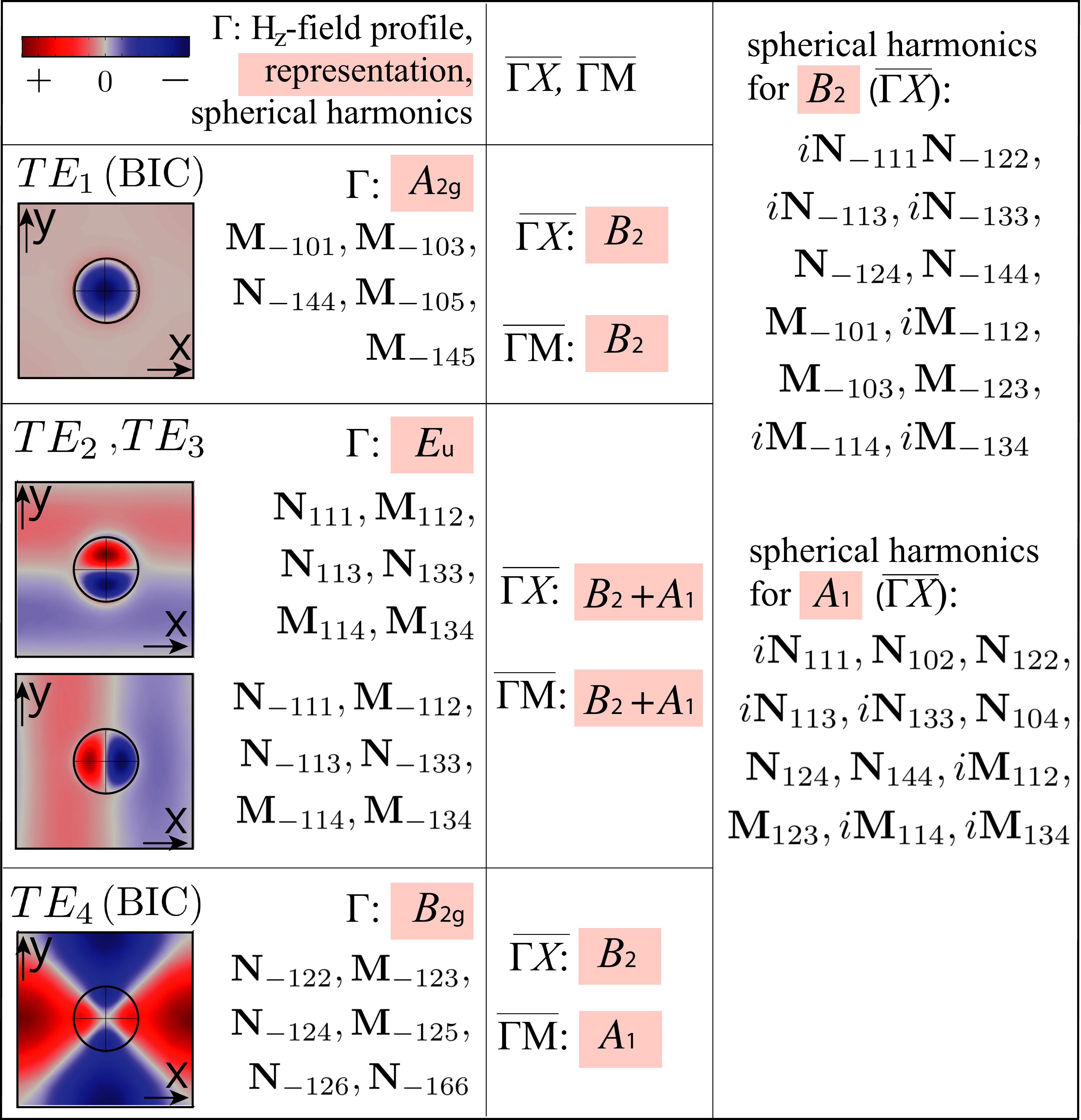}
   \label{tableTE}
\end{table}

%========================Table ends============================================================

%The aim of this section is to reveal which of the coefficients $D_{p_ip_rmn}$ are non-zero for each mode, in accordance with the mode symmetry imposed by the symmetry of the periodic structure \cite{multsym1, multsym}.

The  group-theoretic approach is a powerful method which is widely used for analyzing the properties of periodic photonic system~\cite{group1, group2, group3, Sakoda, groupTuz}. In this section, we  apply this method to reveal which of the coefficients $D_{p_ip_rmn}$ are non-zero in accordance with the mode symmetry imposed by the symmetry of the periodic structure \cite{multsym1, multsym} and to explain the formation of  BIC. To make further analysis more illustrative we will provide it by the example of a square periodic array of dielectric spheres shown in the inset in Fig.~\ref{band}(a). Nevertheless, it is necessary to highlight that further considerations remain true for any kind of structures with square lattice with the point group symmetry $D_{4h}$.

Figure~\ref{band}(a) shows the dispersion of eigenmodes  along the  $\overline{\Gamma X}$ direction for a square periodic array of dielectric spheres with permittivity $\veps_2=12$ embedded in the air with permittivity $\veps_1=1$. The dispersion is calculated numerically using  COMSOL Multiphysics package.  The radius of the spheres is $a=100$ nm and the period of the array is  $d=600$ nm. By the analogy with a dielectric slab waveguide, the eigenmodes of the 2D array are split into transverse electric (TE) and transverse magnetic modes (TM), which have mainly  the $z$-component of magnetic and electric field, correspondingly. Figure~\ref{band}(b) shows the dependence of the Q-factor on the Bloch wavenumber $k_x$ for the modes, which are BICs at the $\Gamma$-point.  One can see that additionally to at-$\Gamma$-BIC, TM$_3$ mode turns into  off-BIC in the middle of the Brillouin band (accidental BIC).

%The divergence of the quality factor for $TM_3$ mode (Fig.~\ref{band} b))  evidences the formation of  a bound state. The explanation of the observed effect is provided in the following section  in terms of MDM by considering  a two-dimensional array with an arbitrary unit cell.

%========================Table============================================================

 \begin{table}
 	 \caption{Multipolar content and irreducible representations  of TM-modes. The $z$-component of the electric field $E_z$ is shown in the inserts. Note that representations can be obtained by replacing for TE-modes $A_2 \leftrightarrow A_1, B_2 \leftrightarrow B_1, u \leftrightarrow g$, and multipolar content can be obtained by replacing   $p_r \leftrightarrow -p_r, \vec M \leftrightarrow \vec N$ }
	\includegraphics[width=0.99\linewidth]{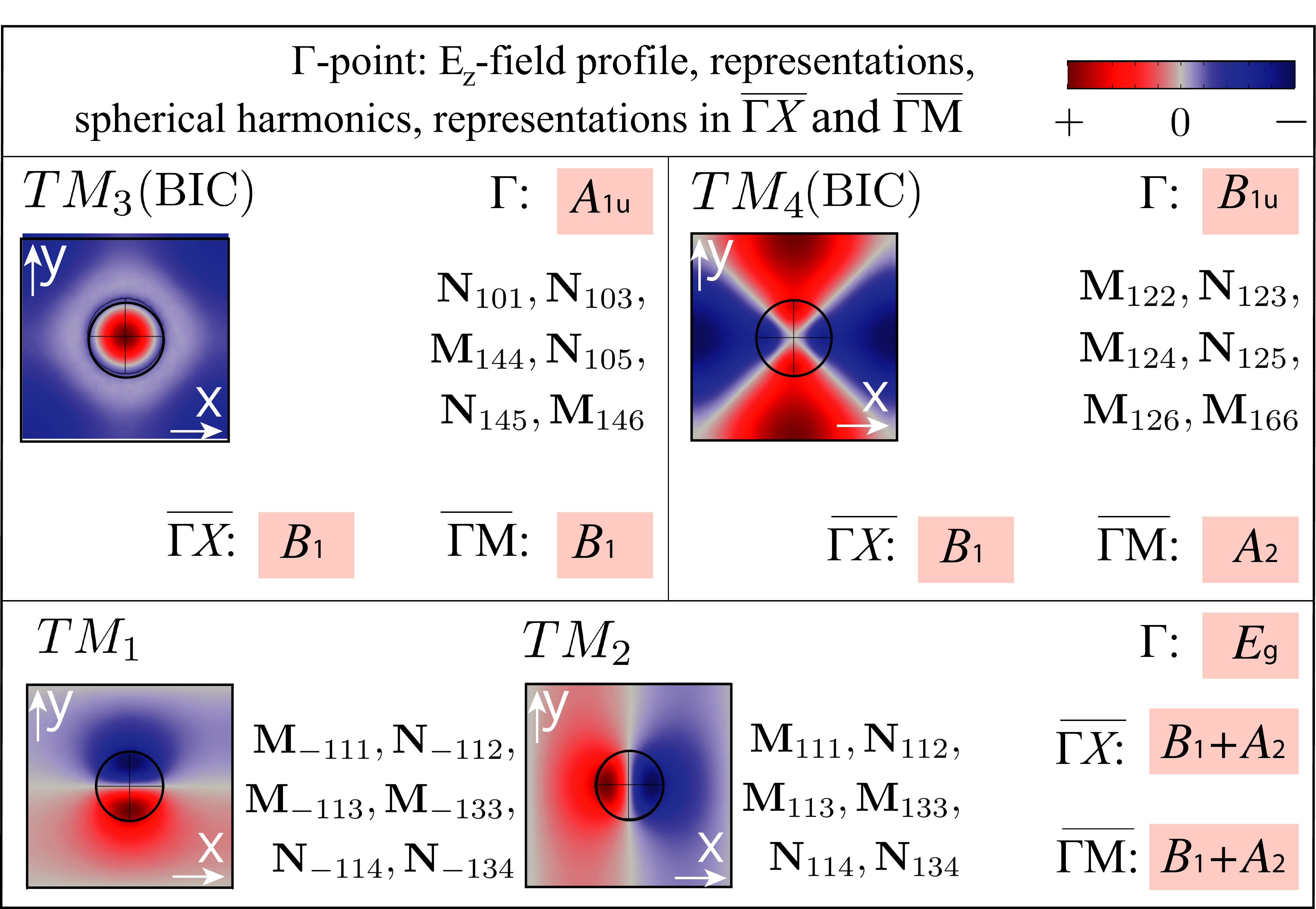}
 \label{tableTM}
\end{table}
%========================Table ends============================================================

The eigenmodes of periodic structures have certain symmetry in accordance with the fact that every mode is transformed  by an irreducible representation of the structure's symmetry group \cite{agranovich2013crystal, Ivchenko1995}. While Bloch functions are already the basis functions of the translation group irreducible representation labeled by $\vec k_b$, under point group operations the basis functions with different $\vec k_b$ transform through each other. They carry high-dimensional representations of the space group. We are interested in the multipolar content of the mode with particular $\vec k_b$, so we look at the point group of $\vec k_b$, i.e. the subgroup of the whole point group, which transforms the $\vec k_b$ into an equivalent. Note that symmetry of the unit cell should be also taken into account since it alters the point group of the full structure.

At the $\Gamma$-point, the group of $\vec k_b$ is the whole group $D_{4h}$,  $\vec k_b=0$ and it is not transformed. At $\overline{\Gamma X}$-valleys the point group of $\vec k_b$ is  $C_{2v}$, which consists of $\pi$-rotation around the $x$- or $y$- axis and two plane reflections at $z=0$ and at $y=0$ or $x=0$. These operations keep vector $\vec k_b$ invariant. Analogously, in the $\overline{\Gamma M}$-valley the group is also $C_{2v}$. Solutions with particular $\vec k_b$ are transformed by one of the $\vec k_b$-group irreducible representations.
Thus, since the solution is transformed as a basis function of some particular representation, the multipolar content is strictly limited. Namely, all the multipoles with non-zero contribution must be transformed by the similar irreducible representation of the $\vec k_b$-group. {We will use common notations for irreducible representations, listed, for example, in Ref.~\onlinecite{jpgoss}.}

{\it $\Gamma$-point}.  For example, we consider the TE$_1$ mode of the square array, which is transformed by $A_{2g}$ at the $\Gamma$-point. Under the transformations of $D_{4h}$ group the only low-order multipoles transformed by $A_{2g}$ are magnetic dipole $\vec M_{-101}$, magnetic octupole $\vec M_{-103}$, and electric hexadecapole  $\vec N_{-144}$.  All of them are invariant under $C_4$ rotations, even under inversion and $z=0$-plane reflection and odd under other $D_{4h}$ transformations.  Higher-order multipoles which behave in the same way are also presented in the multipolar content of this mode.

Analogously, we classify all possible multipoles at the $\Gamma$-point in accordance with their symmetry and provide the tables with multipolar content of the modes (Tables~\ref{tableTE} and ~\ref{tableTM}). Note, that TE$_2$ and TE$_3$ modes degenerate, and they transform through each other as two basis functions of the representation $E_u$.

 %TE mode is even under $z\rightarrow -z$ reflection and TM is odd.
%The eigenfunctions in the high-symmetry points or valleys are transformed under any representation of  the point group, which operations transform k-vector into equivalent or the same. These functions are the basis of similar representation of the translation group.

%Using the method from [Sakoda pp. 51-53, 182] we find that in $\Gamma_1$-point the considered four TE modes are transformed by $A_{2g},  B_{2g}, E_u$ irredusible representations of  $D_{4h}$. The four TM modes in $\Gamma_1$-point are transformed by $A_{1u}, B_{1u}, E_g$. $E_g$ and $E_u$ representations are two-dimensional, so we have doubly-degenerated modes. The only possible multipoles which can be part of these modes must be transformed under the same representation of $D_{4h}$  as the mode.
%With help of [ $http://gernot-katzers-spice-pages.com/character_tables/D4h.html $] we find these multipoles and \red{make table later}
%$$TE_1 \ \ A_{2g} - \vec M_{-101}, \vec M_{-103}, \vec N_{-144}, \dots$$
%$$TE_4 \ \ B_{2g} - \vec N_{-122}, \vec M_{-123}, \vec N_{-124}, \dots$$
%$$TE_2 \ \  E_u 1 - \vec N_{-111}, \vec M_{-112}, \vec N_{-113}, \vec N_{-133}, \dots$$
%$$TE_3 \ \  E_u 2 - \vec N_{111}, \vec M_{112},\vec N_{113}, \vec N_{133}, \dots$$
%$$TM_1 \ \ A_{1u} - \vec N_{101}, \vec N_{103}, \vec M_{144}, \dots$$
%$$TM_4 \ \  B_{1u} - \vec M_{122}, \vec N_{123}, \vec M_{124}, \dots$$
%$$TM_2 \ \ E_g 1 - \vec M_{-111}, \vec N_{-112}, \vec M_{-113}, \vec M_{-133}, \dots$$
%$$TM_3 \ \ E_g 2 - \vec M_{111}, \vec N_{112},\vec M_{113}, \vec M_{133}, \dots$$

{{\it $\overline{\Gamma X}$ valley}. After the symmetry reduction, when we step out the $\Gamma$-point into $\overline{\Gamma X}$ valley, some symmetry operations remain, e.g. mirror reflections in $z=0$ and $y=0$ planes and rotation by $\pi$ around the $x$-axis. Eigenmodes must behave in the same way under these symmetry operations, as at the $\Gamma$-point.

As an example, we consider TE$_1$ mode, which transforms by $A_{2g}$ at the $\Gamma$-point, and TM$_3$ mode ($A_{1u}$) at the $\overline{\Gamma X}$ valley.
Using the compatibility relations \cite{Sakoda}, we obtain that the mode which transforms by $A_{2g}$ at the $\Gamma$-point is transformed under $B_2$ representation of the group $C_{2v}$ in the $\overline{\Gamma X}$ valley. The TM$_3$-mode, which is $A_{1u}$ at the $\Gamma$, is transformed under $B_1$ in $\overline{\Gamma X}$. }%This means that after the symmetry reduction when we step out the $\Gamma$-point into $\overline{\Gamma X}$ valley, some symmetry operations remain, e.g. mirror reflections in $z=0$ and $y=0$ planes and rotation by $\pi$ around the $x$-axis. The eigenmodes must behave in the same way under these symmetry operations, as in the $\Gamma$-point.} 
{For  the $A_{2g}$ mode} at {the} $\Gamma$-point, which is odd under reflection in the $y=0$ plane and $\pi$-rotation around $x$ and even under reflection in $z=0$ plane, the only possible multipoles in $\overline{\Gamma X}$ valley should have the same symmetry properties.
For the $A_{1u}$ mode the possible multipoles in the $\overline{\Gamma X}$ valley must be odd under reflection in the $z=0$ plane and $\pi$-rotation and even under reflection in the $x=0$ plane. Low-order possible multipoles in the $\overline{\Gamma X}$ valley are listed at the right column %bottom
 of Table~\ref{tableTE} {for TE modes, and for TM modes are easily derived by replacing $A_2 \leftrightarrow A_1, B_2 \leftrightarrow B_1$,  $u \leftrightarrow g, $\   $p_r \leftrightarrow -p_r, \ \vec M \leftrightarrow \vec N$.}
Analogously, for the  $\overline{\Gamma M}$ valley we have specific symmetry in the $x=y$ or $x=-y$ direction  and possible multipoles are the same as those which are  transformed by the same representation in $\overline{\Gamma X}$ valley, but rotated by $\pi/4$ with help of Wigner D-matrixes \cite{Theorem2008, Aubert2013}.

%Some of them are listed below:

 %Analogously, for the  $\overline{\Gamma M}$ valley we have specific symmetry under reflection in $x=y$ plane? and possible multipoles are listed below.
 %$$i\vec N_{o11}-i\vec N_{e11},  \vec N_{e22}, i\vec N_{o13}-i\vec N_{e13}, i\vec N_{o33}-i\vec N_{e33}, \vec N_{e24}, \vec N_{o44},  $$
 %$$\vec M_{e01}, i \vec M_{e12}-i\vec M_{o12}, \vec M_{e03}, \vec M_{o23},i \vec M_{e14}-i\vec M_{o14}, i\vec M_{e34}- i\vec M_{o34} $$

 %For the TM $A_{1u}$ mode the possible multipoles in the $\overline{\Gamma X}$ valley bust be odd in the $z=0$ plane reflection and $\pi$-rotation and even in the $x=0$ plane reflection. Some of them are listed below (\red {table?})
%$$ i\vec N_{101},  \vec N_{112},i \vec N_{103},  \vec N_{114}, \vec N_{134},$$
%$$ \vec M_{111}, i\vec M_{122}, \vec M_{113}, \vec M_{133},  i\vec M_{124},  i\vec M_{144}$$

\section{Multipolar composition of the eigenmodes in periodic structures} \label{comss}

\subsection{Multipole analysis of metasurfaces and photonic crystal slabs}

Before applying the developed approach to certain structures, we would like to emphasize that the power of the group-theoretic approach and multipole decomposition method is that these methods applied equally for metasurfaces (arrays of meta-atoms) and photonics crystal slabs (slabs with holes). For example, a square array of dielectric spheres and dielectric slab with circular holes arranged in a square lattice are identical from the point of view of the group theory. Therefore, these structures have the same classification of eigenmodes and, that more important, the same multipole content (Mie-resonances). The difference will be only in amplitudes of the multipole coefficients.

At first glance it may seem unphysical to characterize the unit cell with a hole by Mie resonances. This contradiction can be eliminated by means of the proper choice of the unit cell. Indeed, Fig.~\ref{correspondece} shows the modes of the photonic-crystal slab and metasurface  corresponding to the same irreducible representation $B_\text{1u}$ (see Table~\ref{tableTM}). One can see that the mode structures are completely identical.  However it is more physical to assign Mie resonances not the unit cell with a hole but to the cross-like shape unit cell (see the low panel in   Fig.~\ref{correspondece}). In particular, it was shown in \cite{QuanlongYang2019} that the optical response of both photonic-crystal slab and metasurface is dominated by the electric and magnetic Mie-type dipole resonances, and both type of the structures provides $2\pi$ phase control of light.

\begin{figure}
	\includegraphics[width=0.95\linewidth]{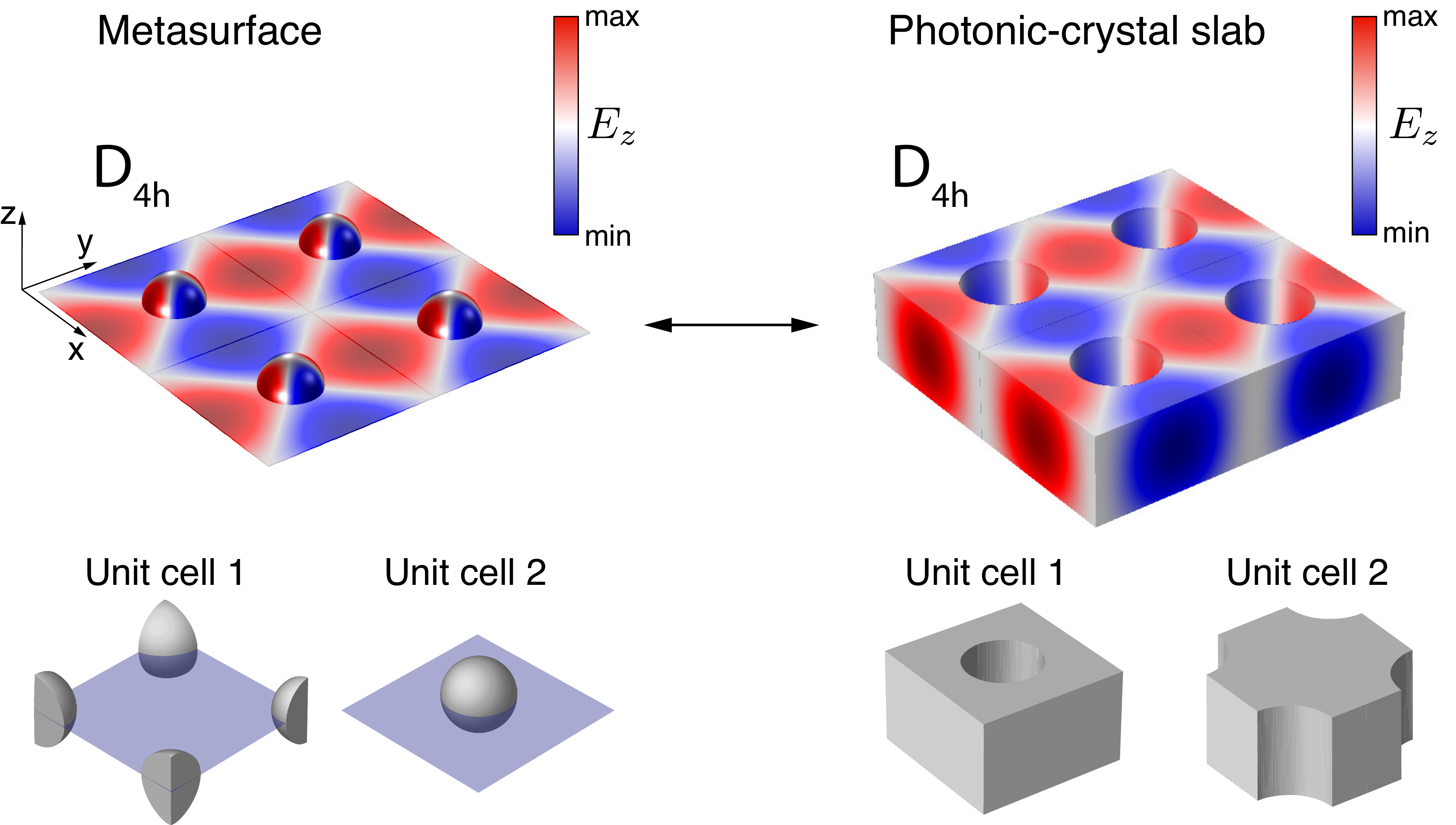}
	 \caption{Distribution of $E_z$ component of electric field of the modes corresponding to the same irreducible representation $B_\text{1u}$ (see Table~\ref{tableTM}) in the photonic-crystal slab and metasurface.} \label{correspondece}
\end{figure}

\subsection{Periodic arrays of dielectric spheres}

 \begin{figure}
	\includegraphics[width=0.95\linewidth]{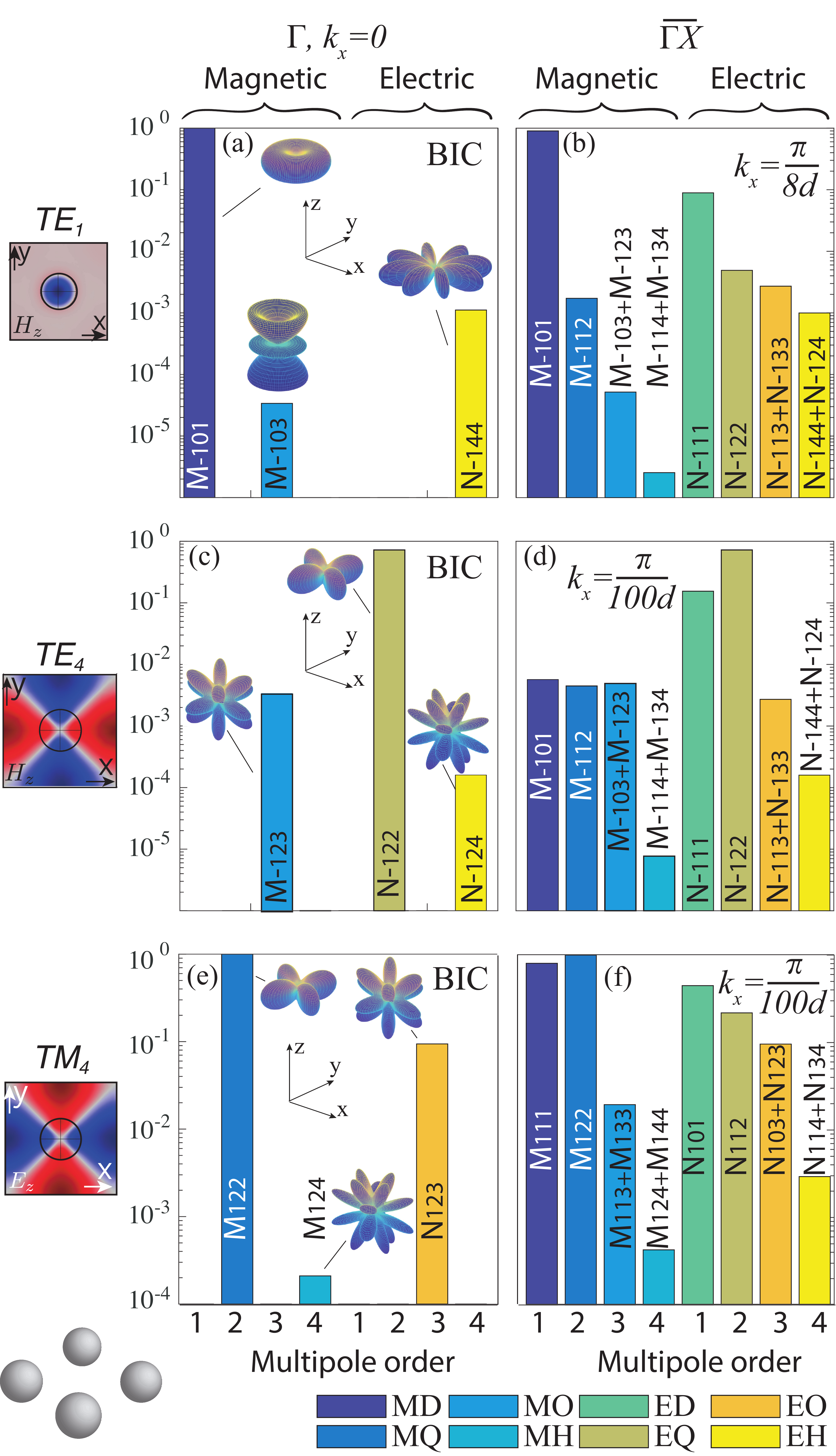}
	%\includegraphics[width=0.95\linewidth]{probe6-eps-converted-to.pdf}
	 %\caption{Appearance of the multipoles in the $\overline{\Gamma X}$-valley due to lowering of the symmetry. \red {Vertical axis shows values of normalized $\tilde D$ coefficients.} (a) For the TE$_1$ mode only MD, MO and EH are presented at the $\Gamma$-point.  (b) In  the $\overline{\Gamma X}$-valley multipoles of all orders are presented. (c) TE$_4$ and (e) TM$_4$ modes at the $\Gamma$-point. Only multipoles with $m=2$ are presented if $n \leq 4$.  (d) TE$_4$ and (f) TM$_4$ modes are close to the light line, so we can only move by $\pi/100d$ but this is still enough for all contributions to appear.  D-dipole $(n=1)$, Q-quadrupole $(n=2)$, O-octupole $(n=3)$, H-hexadecapole $(n=4)$. We also give far-field diagrams of the multipoles at the $\Gamma$-point.}
	 \caption{Multipolar content of $TE_1$ (a,b) , $TE_4$ (c,d) and $TM_4$ (e,f) modes of the metasurface. Vertical axis shows values of normalized coefficients $\tilde D$. Horizontal axis shows multipolar order $n$. The letters in legend encodes the multipoles as follows:  $ED\,(MD)$ - electric (magnetic) dipole ($n=1$), $EQ\,(MQ)$ -electric (magnetic) quadrupole ($n = 2$), $EO\,(MO)$ - electric (magnetic) octupole ($n = 3$), $EH\,(MH)$ - electric (magnetic) hexadecapole ($n = 4$).}
	 \label{figmult}
\end{figure}

 In order to understand how the interference of the multipole moments forms the BICs, we applied a method of multipole decomposition by expanding the eigenmodes fields inside the nanoparticles in terms of spherical harmonics~\cite{Jackson1965, Grahn2012}. Figure~\ref{figmult} shows the numerical results for the multipole decomposition for TE$_1$,  TE$_4$,  and TM$_4$ modes at the $\Gamma$-point and in a point of the $\Gamma$X-valley. At the high-symmetry $\Gamma$-point only a small fraction of all possible multipoles is presented, while after lowering the symmetry extra multipoles appear out of the $\Gamma$-point. Figure~\ref{figmult2} shows the numerical results for the multipole decomposition for TM$_3$ modes at the $\Gamma$-point, in a point of the $\Gamma$X-valley, and at the off-$\Gamma$ BIC.

{\it At-$\Gamma$-BIC.}  One can see, that  for at-$\Gamma$ TE$_1(A_{2g}$) BIC-mode magnetic dipole along the $z$-axis ($\vec M_{-101}$)  is mainly contributed  [Fig.~\ref{figmult}(a)]. Its directivity pattern restricts radiation along the $z$-axis, and the radiation patterns of all remaining multipoles at the $\Gamma$-point are also zero along the $z$-axis. Other directions of radiation are forbidden due to the subdiffraction regime. Similarly, at-$\Gamma$ TM$_3$($A_{1u}$) BIC is formed by electric dipole along the $z$-axis ($\vec N_{101}$) mostly, prohibiting the radiation in vertical direction itself  [Fig.~\ref{figmult2}(a)].  {The TE$_4$ ($B_{2g}$) and  TM$_4$ ($B_{1u}$) modes' lowest multipoles are the electric quadrupole $\vec N_{-122}$  and the magnetic quadrupole $\vec M_{122}$ respectively. They have $m=2$, which also prohibits the radiation} [Figs.~\ref{figmult}(c) and \ref{figmult}(e)]. {In contrast to BICs, the radiative modes ($E_g, E_u$) degenerate at the $\Gamma$ point since they are transformed by the two-dimensional representations. From the symmetry-group approach, we know that TE$_{2,3}$ and TM$_{1,2}$  modes contain electric $\vec N_{\pm111}$ and magnetic $\vec M_{\pm111}$ spherical harmonics. The  numerical multipole expansion shows that degenerated modes contain in-plane electric or magnetic dipole moments as the main contribution. %\sout{Both numerical simulation and a symmetry-group approach validate that there are no other modes in the system with square lattice. }{\bf there are some other types of modes with different symmetry))}
 {These numerical results validate the symmetry-group approach for the system with square lattice,  confirming that any symmetry-protected BIC is characterized by multipole moments with $m\neq 1$.}

 {\it ~Off-$\Gamma$ BIC.} Away from the $\Gamma$-point other multipoles appear in decomposition [Figs.~\ref{figmult}(b),~\ref{figmult}(d),~\ref{figmult}(f) and ~\ref{figmult2}(b)] and BIC is destroyed turning into a resonance state.{ As mentioned earlier in the Section \ref{formulae}, the accidental off-$\Gamma$ BICs is formed due to  either in- or anti-phase   contributions of different multipoles.} We obtain the same result for TM$_3$ mode with off-$\Gamma$ BIC in the $\Gamma$X-valley [Fig.~\ref{figmult2}(c)]. This mode is transformed by $B_1$ representation and consist of the multipoles, which are odd under reflection in the $z=0$ plane and $\pi$-rotation around $x$-axis and even under $y=0$ reflection, eg. $i\vec N_{101}, \vec N_{112}, i\vec N_{103}, i\vec N_{123},  \vec M_{111}, i\vec M_{122}, \vec M_{113}, \vec M_{133}$. They sum up into $TM$-polarized wave in the direction given by vector $\vec k_1$ and cancel each other in the off-$\Gamma$ point forming accidental BIC, shown in Fig.~\ref{band}(b).
In addition to all, it is well known, that the $z=0$ plane reflection symmetry of the structure is required to obtain the off-$\Gamma$ BIC~\cite{Hsu2013}. Indeed, in lack of such symmetry, each mode would contain both odd and even multipoles under reflection in  the $z=0$ plane. To restrict the radiation both in the upper- and lower half-spaces, odd and even multipoles should be summed up into zero independently, while for the symmetric structure only one type of multipoles is presented for each mode, which makes it possible to achieve the BIC by tuning the structure parameters. %\sout{which is harder to achieve by tuning the structure parameters.}\commentMisha{it is not very clear to me. "hard" does not mean impossible}

\subsection{Photonic crystal slab}

{We extend our approach beyond the 2D array of spheres and apply it  to the photonic crystal slab of the same symmetry. We consider a dielectric slab with a square array of cylindrical holes studied in~\cite{Hsu2013}. Both at-$\Gamma$ (symmetry-protected) and off-$\Gamma$ (accidental) BICs appear in the lowest TM band referred to as TM$_h$. This mode has the  field profile of the same symmetry as the TM$_4$ mode of the array of the spheres and is transformed by $B_{1u}$ representation.

 \begin{figure}
	\includegraphics[width=0.99\linewidth]{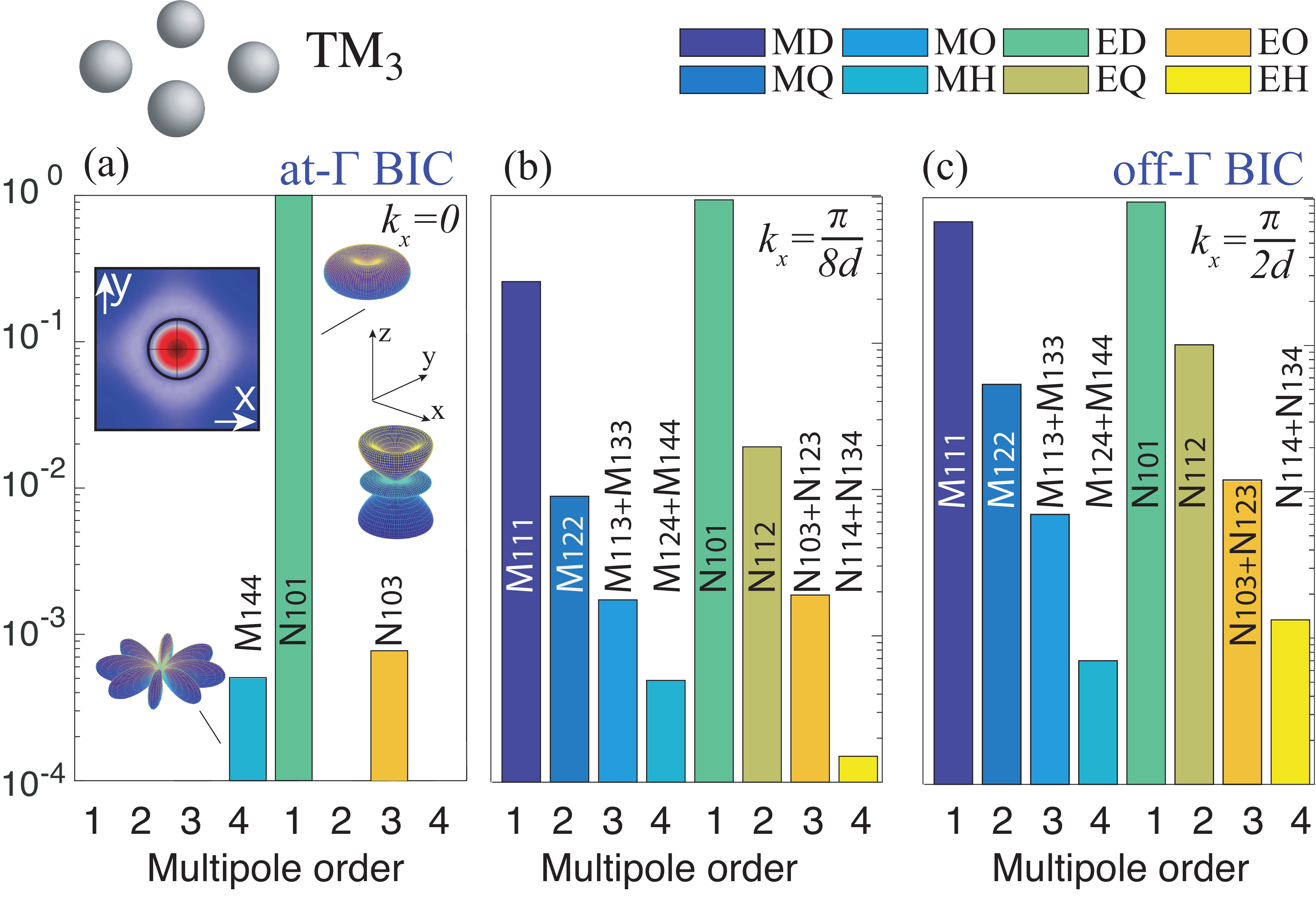}
	% \caption{Appearance of the multipoles in the $\overline{\Gamma X}$-valley due to lowering of the symmetry for the TM$_3$ mode. The vertical axis shows values of normalized $\tilde D$ coefficients. (a) Multipolar content of the at-$\Gamma$ BIC.  At the $\Gamma$-point only ED, EO and MH multipoles are presented, far-field diagrams of the multipoles are given. (b) Multipolar content in the $\overline{\Gamma X}$-valley.  Multipoles of all orders are presented. (c) Multipolar content of the off-$\Gamma$ BIC is the same as in the rest of the $\overline{\Gamma X}$-valley, because lowering of the symmetry doesn't take place. BIC appears due to destructive interference of the radiation patterns of all multipoles. This is possible because all multipoles are in phase.  D-dipole $(n=1)$, Q-quadrupole $(n=2)$, O-octupole $(n=3)$, H-hexadecapole $(n=4)$.}
	 \caption{ Multipolar content of $TM_3$ mode of the metasurface for three $k_x$ Bloch vectors corresponding to the symmetry-protected BIC(a), leaky mode (b) and accidental BIC (c). Vertical axis shows values of normalized $\tilde D$ coefficients, horizontal axis shows multipolar order $n$. The letters in legend encodes the multipoles as follows:  $ED\,(MD)$ - electric (magnetic) dipole ($n=1$), $EQ\,(MQ)$ -electric (magnetic) quadrupole ($n = 2$), $EO\,(MO)$ - electric (magnetic) octupole ($n = 3$), $EH\,(MH)$ - electric (magnetic) hexadecapole ($n = 4$).}
	 \label{figmult2}
\end{figure}

\begin{figure}
	\includegraphics[width=0.95\linewidth]{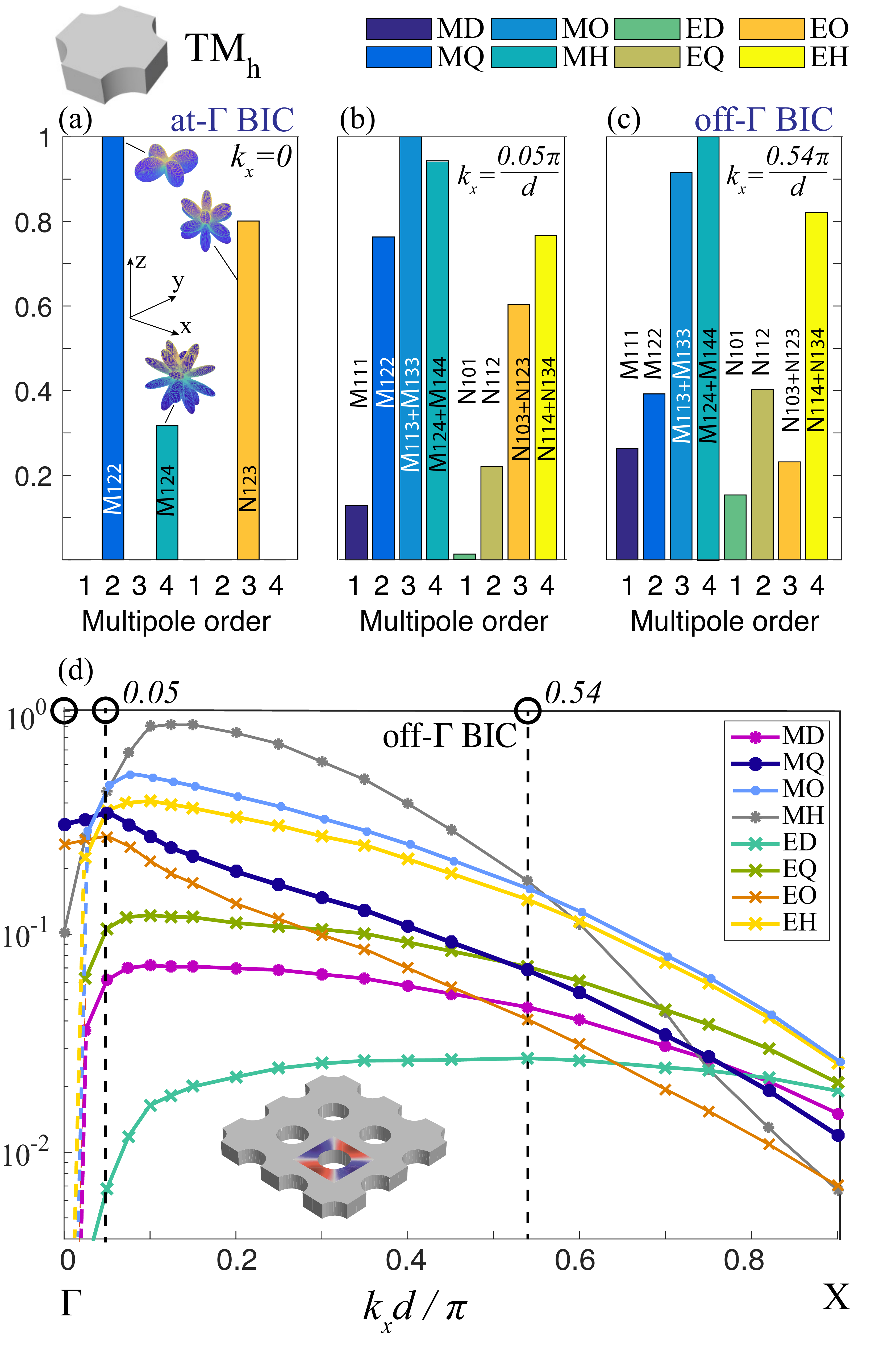}
	 %\caption{Appearance of the multipoles in the $\overline{\Gamma X}$-valley due to lowering of the symmetry  for the TM$_h$-mode of the slab~\cite{Hsu2013} with $B_{1u}$ representation. \red{Vertical axis shows values of normalized $\tilde D$ coefficients.}  (a) Multipolar content of the at-$\Gamma$ BIC.  At the $\Gamma$-point only MQ, MH, and EO multipoles are presented, far-field diagrams of the multipoles are given. (b) Multipolar content in the $\overline{\Gamma X}$-valley.  Multipoles of all orders are presented. (c) Multipolar content of the off-$\Gamma$ BIC is the same as in the rest of the $\overline{\Gamma X}$-valley, because lowering of the symmetry doesn't take place. (d) Dependence of the multipolar contributions from the $k_x$ at the $\overline{\Gamma X}$-valley.}
	 \caption{Multipolar content of $TM_h$ mode of the PhC slab for three $k_x$ Bloch vectors corresponding to the symmetry-protected BIC(a), leaky mode (b) and accidental BIC (c). Vertical axis shows values of normalized $\tilde D$ coefficients, horizontal axis shows multipolar order $n$. Figure (d) depicts how the multipolar contributions change along $\overline{\Gamma X}$ valley. The letters in legend encodes the multipoles as follows:  $ED\,(MD)$ - electric (magnetic) dipole ($n=1$), $EQ\,(MQ)$ -electric (magnetic) quadrupole ($n = 2$), $EO\,(MO)$ - electric (magnetic) octupole ($n = 3$), $EH\,(MH)$ - electric (magnetic) hexadecapole ($n = 4$).}
	 \label{figmult22}
\end{figure}

The description of the far field defined by the Eq.~\eqref{eq:farfield} is still valid, but {the} coefficients $\tilde D_{p_ip_rmn}$ in general cannot be expressed analytically through $D_{p_ip_rmn}$ (see Appendix \ref{appcoef}). However, only the multipoles which are presented in the field inside the slab contribute into far field,  $\tilde  D_{p_ip_rmn}=0$ if $D_{p_ip_rmn}=0$. The multipolar content for the considered slab  is the same as in the periodic array of spheres because the modes of the slab have the same symmetry as the modes of the array. The multipole decomposition of TM$_h$ mode [Fig.~\ref{figmult22}(a)] reveals that a magnetic quadrupole $\vec M_{122}$ and electric octupole $\vec N_{123}$ make the major contribution to the at-$\Gamma$ BIC, as well as to the TM$_4$ mode of the array of the spheres.  However, the $B_{1u}$ mode of the photonic crystal slab is the lowest-energy TM mode while for the array of spheres it has the highest energy among modes under diffraction limit. %\sout{This is due to the field distribution in this mode, and variational principle~\cite{Joannopoulos2008}. Namely}
{Due to the variational principle~\cite{Joannopoulos2008}, for the mode of such symmetry, the electric field is more concentrated inside the high-index material in case of photonic crystal slab, minimizing the energy of the mode.}
Although dispersion curves $\omega (k)$ of the modes TM$_4$ and TM$_h$ behave completely differently, multipole decomposition proves a common origin of them. At the $\Gamma$-point, we obtain contributions of multipoles only with {$m=2, 6, 10,$ etc.} for both modes, and none of these multipoles contribute in the far-field. At the $\overline{\Gamma X}$-valley, the multipolar content of the TM$_4$ and TM$_h$ is similar [Figs.~\ref{figmult}(e), \ref{figmult}(f),~\ref{figmult22}(a), and~\ref{figmult22}(b)]. However, for the TM$_h$ it is possible  to obtain an accidental off-$\Gamma$ BIC  in the $\overline{\Gamma X}$-valley [Fig.~\ref{figmult22}(c)].  Away from the $\Gamma$-point, multipole contributions change smoothly keeping the multipolar content invariable, and at a particular wave vector $ k_x$ multipoles interfere destructively forming the accidental BIC.

%%%%%%%%%%%%%%%%%%%

 %Let us consider the TM$_h$-mode of the slab with $B_{1u}$ representation. We built the multipolar contributions for different points of $\Gamma$X-valley in the Fig.~\ref{mult_hole},~\ref{figmult22}. . \red{write something}

%\begin{figure}
%	\includegraphics[width=0.7\linewidth]{mult_holes1-eps-converted-to.pdf}
%	 \caption{Multipolar decomposition of the TM$_{h}$-mode. Contributions of electric (E) and magnetic (M) spherical multipoles, which are normalized $D$-coefficients of %decomposition of electromagnetic field of photonic crystal slab in the $\overline{\Gamma X}$-valley, given by \eqref{ein}} \label{mult_hole}.
%\end{figure}

%To explain the BICs formation deeply, we apply a group-theoretical approach to the mode analysis in the next section. We also discuss the multipolar content of
%By eigenmode analysis, we known that at a particular wave vector $k_x$ the accidental BIC emerges.
%that one cannot define appearance of the BIC in the $\overline{\Gamma X}$-valley relying on the multipole content only as Fig.~\ref{fig4}.
% both periodic array of the spheres and photonic crystal and find, that multipole contents of the modes
%the symmetry protected
%BICs and provide symmetry classification and multipolar content of presented modes.

%====================================FIGURE==================================================================

%\section{Discussions}
%\label{Sec:Discussion}

%=======================TABLE=====================================================================

%-----------------------------------------------------%
\section{Summary and outlook}

Importantly, our approach based on the multipole decomposition analysis of individual meta-atoms not only explains clearly and in simple physical terms the origin of both symmetry protected and accidental bound states in the continuum, but it has also a prediction power and may be employed for both prediction and engineering different types of BICs. As an example, we consider a metasurface consisting of meta-atoms packed in a subwavelength 2D lattice, which are polarized purely as octupoles, for example, $\vec M_{-103}$ (see Fig.~\ref{fig:example}). Each octupole of this type has a nodal cone and, therefore, we can expect that in a periodic subwavelength array of such meta-atoms, BICs form a line in the reciprocal space. However, to observe this phenomenon, the effective polarizability of the unit cell accounting for the interaction between all meta-atoms should not depend on the Bloch wavenumber or have very weak dependence. In other words, the line of BIC could be observed in metasurfaces  with suppressed spatial dispersion that is still a challenging problem. Interestingly, such a kind of BIC will be observed for the same directions independently on the lattice symmetry of the metasurface.

\begin{figure}[t]
   \centering
   \includegraphics[width=0.85\linewidth]{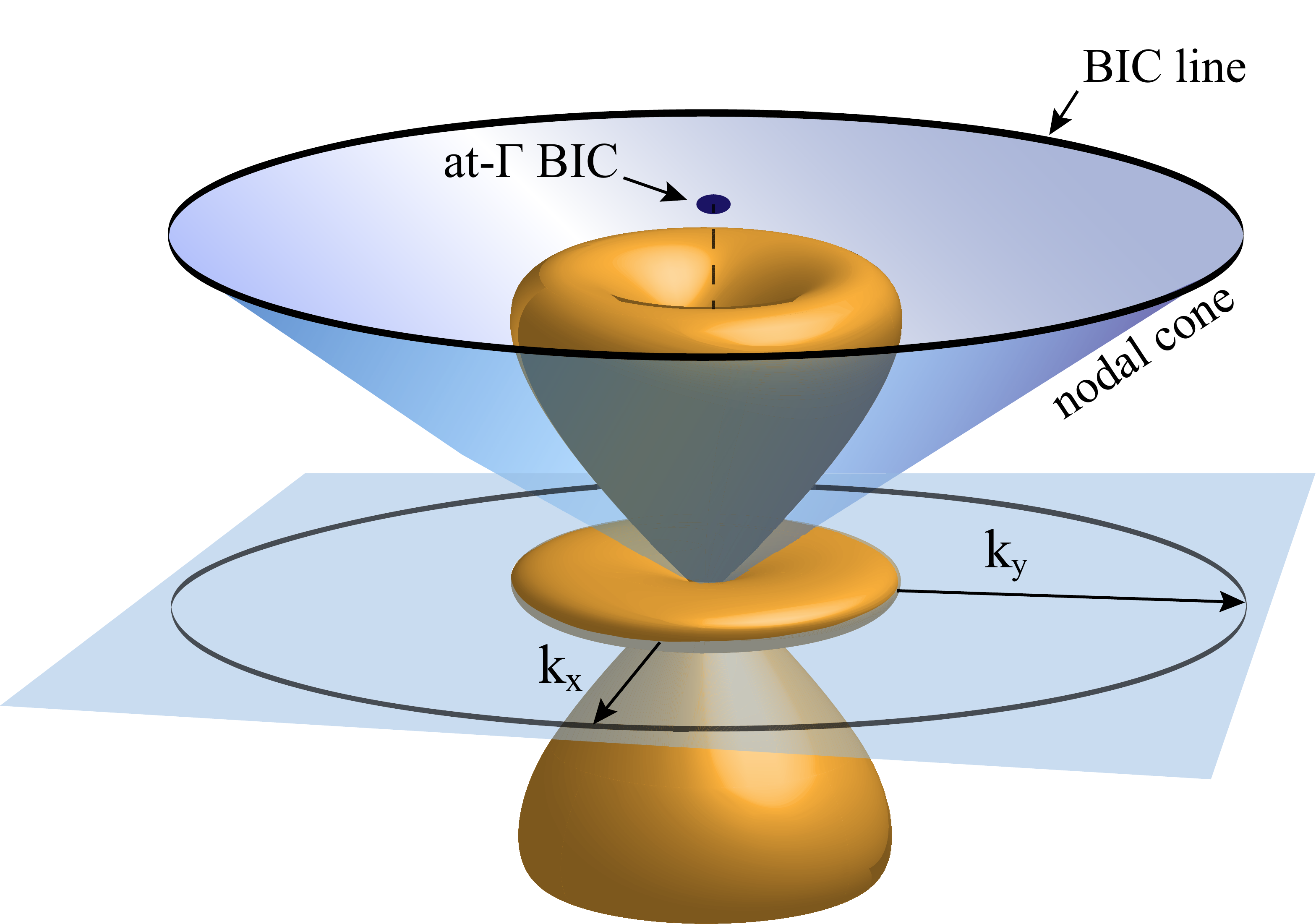} % requires the graphicx package
   \caption{BIC line formed by metasurface composed of point octupoles with radiation pattern $\vec N_{103}$.}
   \label{fig:example}
\end{figure}

{We can expand this discussion even further, by adding $\vec M_{-101}$ dipole mode into consideration. To obtain the Q-factor of the at-$\Gamma$ BIC, we should find the  dependence of energy loss rate on $\vec k-$vector in Eq.~\eqref{eq:farfield}, assuming stored energy is almost constant when $\theta_k \rightarrow 0$. The asymptotic behavior for $\theta_k \rightarrow 0$ of $\vec Y_{p_ip_rmn}\left(\frac{\vec k}{|\vec k|}\right)$ functions is defined by $\frac{d P_n^m( \cos\theta_k)}{d \theta_k}$. So, $Y_{1-103}\left(\frac{\vec k}{|\vec k|}\right)\propto (-6k_x+17k_x^3/2)  \vec e_{\varphi}$, and $Y_{1-101}\left(\frac{\vec k}{|\vec k|}\right) \propto (k_x-k_x^3/6) \vec e_{\varphi}$.  We can manage their relative contribution, and at particular point when coefficient before dipole is six times larger, we obtain that linear terms cancel each other and field asymptotic is proportional to $k_x^3$, so the quality-factor growth is proportional to $k_x^{-6}$. Similar effect was observed in \cite{qfactorarxiv} for the photonic crystal slab. However, considering realistic situations, we should take into account all possible multipolar contibutions, including terms with $m=1$, where the multipole contribution growth rate plays role but not the asymptotic behavior.    }Thus, the multipole origin of BIC makes a new query for metasurfaces with a suppressed spatial dispersion.

In summary, we have demonstrated that symmetry-protected bound states in the continuum in dielectric metasurfaces and photonic crystal slabs at the frequencies below the diffraction limit are associated with the multipole moments of the elementary meta-atoms which do not radiate in the transverse direction. For any type of metasurfaces, the symmetry-protected bound states in the continuum can be observed only if there exist no multipoles with the azimuthal index $m=1$ in the multipole decomposition. The symmetry approach allows to determine which multipolar content the lattice eigenmodes have, and it can be analogously applied to the structures with different symmetries, for example, hexagonal lattices or arrays of nanoparticles of an arbitrary shape and in-plane broken symmetry. Similarly, we have revealed that the accidental bound state in the continuum corresponding to an off-$\Gamma$ point in the reciprocal space is formed due to destructive interference of the multipole fields in the far-zone. We have provided the general tools for the analysis of bound states in the continuum based on the irreducible representation of the appropriate photonic band. We believe that our results will provide a new way for designing high-quality resonant photonic systems based on the physics of bound states in the continuum.

\acknowledgements
The authors acknowledge useful discussions with I.D. Avdeev. Y.K. acknowledges a support from
the Strategic Fund of the Australian National University. Z. S. acknowledges support from the Foundation for the Advancement of Theoretical Physics and Mathematics "BASIS" (Russia).

{Z.S. and K.F. contributed equally to this work.}

%-----------------------------------------------------%

%=====================References================================
%\bibliographystyle{acm}
\bibliography{liter_2}

%====================APPENDIX====================================]
\newpage
\appendix

\begin{widetext}

\section{Basic definitions} \label{app:definitions}

 %\begin{flushleft}
% \begin{alignat}{8}
%i\vec N_{111}, \vec N_{102}, \vec N_{122},\\ i\vec N_{113}, i\vec N_{133}, \vec N_{104},\\ \vec N_{124},  \vec N_{144},
  %i\vec M_{112}, \\  \vec M_{123},  i\vec M_{114}, i\vec M_{134} \\
%  i\vec N_{-111} \vec N_{-122},\\
  %i\vec N_{-113}, i\vec N_{-133},\\
  %\vec N_{-124}, \vec N_{-144},  \\
   %\vec M_{-101},  i\vec M_{-112},\\
  %\vec M_{-103}, \vec M_{-123},\\
  %i\vec M_{-114}, i\vec M_{-134},  \\
% \\ i \vec M_{-112}+i\vec M_{112}, \vec M_{-123}, \dots\\%\\ i \vec M_{-114}-i\vec M_{114}, i\vec M_{-134}- i\vec M_{134} \\
%\vec M_{122}, \vec N_{123},\\
% \vec M_{124},  \vec N_{125}, \\
 % \vec M_{126},  \vec M_{166}
  %\end{alignat}
 %  \end{flushleft}

   %$$TE_1 \ \ A_{2g} - \vec M_{-101}, \vec M_{-103}, \vec N_{-144}, \dots$$
%$$TE_4 \ \ B_{2g} - \vec N_{-122}, \vec M_{-123}, \vec N_{-124}, \dots$$
%$$TE_2 \ \  E_u 1 - \vec N_{-111}, \vec M_{-112}, \vec N_{-113}, \vec N_{-133}, \dots$$
%$$TE_3 \ \  E_u 2 - \vec N_{111}, \vec M_{112},\vec N_{113}, \vec N_{133}, \dots$$
%$$TM_1 \ \ A_{1u} - \vec N_{101}, \vec N_{103}, \vec M_{144}, \dots$$
%$$TM_4 \ \  B_{1u} - \vec M_{122}, \vec N_{123}, \vec M_{124}, \dots$$
%$$TM_2 \ \ E_g 1 - \vec M_{-111}, \vec N_{-112}, \vec M_{-113}, \vec M_{-133}, \dots$$
%$$TM_3 \ \ E_g 2 - \vec M_{111}, \vec N_{112},\vec M_{113}, \vec M_{133}, \dots$$

Vector spherical harmonics  are defined as \cite{Bohren}
  \begin{gather}
 \vec{M}_{^{-1}_{\ 1}mn}=\nabla \times (\vec r \psi_{^{\ 1}_{-1}mn})\:, \\
  \vec{N}_{^{\ 1}_{-1}mn}=\frac{\nabla \times \vec{M}_{^{-1}_{\ 1}mn}}{k}\:,
\end{gather}
where
 \begin{gather}
\psi_{1mn}=\cos m\phi P_n^m(\cos\theta)z_n(kr)\:, \\
\psi_{-1mn}=\sin m\phi P_n^m(\cos\theta)z_n(kr)
 \end{gather}
 $z_n(kr)$ can be replaced by spherical bessel function of any kind, and $ P_n^m(\cos\theta)$ are associated Legendre polynomials.

 Vacuum dyadic Green's function expansion in terms of vector spherical harmonics \cite{Tai1972, Yee1994}:

\begin{alignat}{3} \nonumber
 {{\vec{ \hat{G}}_0}}({\vec r, \vec r'})= \frac {i k_1}{4 \pi } \sum_{n=1}^{\infty} \sum_{p_r=\pm 1}\sum_{m=0}^n (2-\delta_0) \frac {2n+1}{n(n+1)} \frac {(n-m)!}{(n+m)!}
\cdot \ \ \ \ \ \ \ \ \ \ \ \  \\
\cdot  \Bigl(  [\vec{M}_{p_rmn}^{(3)}(k_1, \vec r) \otimes{\vec{M}}^{(1)}_{p_rmn}(k_1, \vec r')]
+[{\vec{N}}_{p_rmn}^{(3)}(k_1, \vec r) \otimes{\vec{N}}^{(1)}_{p_rmn}(k_1, \vec r')]\Bigr) \label{gfg} , \ \ \ \  r>r'
\end{alignat}

where superscripts $(1)$ and $(3)$ appear, when we replace $z_n(\rho)$ by spherical Bessel functions, and  the spherical Hankel functions of the first kind, respectively, $\delta_0=1$ when $m=0$, and $\delta_0=0$ when $ m \neq 0$.

Spherical vectors $\vec Y_{p_ip_rmn}$ denote two types of functions, $\vec X_{p_rmn}$ and $\vec Z_{p_rmn}$, defined as \cite{Alaee}
\begin{equation}
\vec X_{-p_rmn}\left(\frac{\vec k}{k}\right)={\nabla \times \left(\vec k Y_{p_rmn}\left(\frac{\vec k}{k}\right)\right)}\
\end{equation} %\frac{\psi_{^e_omn}}{j(kr)}
% \frac{1}{\red{\sqrt{n(n+1)}}}
\begin{equation}
\vec Z_{p_rmn}\left(\frac{\vec k}{k}\right)= { i \frac{\vec k}{k} \times \vec X_{-p_rmn}\left(\frac{\vec k}{k}\right) }\
\end{equation}
where
  \begin{gather}
Y_{1mn}=\cos m\phi P_n^m(\cos\theta) \\
Y_{-1mn}=\sin m\phi P_n^m(\cos\theta)
 \end{gather}
$p_i$ is  $(-1)^{n+1}$ for $\vec X$, and $p_i=(-1)^n$ for $\vec Z$.  Note that the transformation behavior is similar for $\vec W$ and $\vec Y$, $\vec X$ and $\vec M$,  $\vec Z$ and $\vec N$, $\psi$ and $Y$, respectively.

\section{Lattice sums of the spherical harmonics} \label{Weyl}
We assume that multipolar content of the mode is already known, and coefficients in the formula \eqref{ein} are given.
With help of vacuum dyadic Green's function $\hat{  \vec G}_0$, we express the field outside the array
  \begin{alignat}{2} \label{out}
\vec E (\vec r)=\frac{k_1^2}{4\pi}\int d^3 \vec r'' \Delta\varepsilon(\vec r'') \hat {\bf G}_0({\vec r, \vec r''}) \vec E^{in}( \vec r'')= \nonumber \\
=\frac{k_1^2}{4\pi} (\varepsilon_2-\varepsilon_1) \sum_j  \int \limits_{V} d^3  \vec r'   \hat {\bf G}_0({\vec r, \vec r_j+\vec r'}) \vec E^{in}(\vec r_j+ \vec r')
  \end{alignat}
where $k_1 = \sqrt{\varepsilon_1} \frac{\omega}{c} $ is vacuum wavevector, $V$ is the single nanoparticle's volume.

Green's function can be also expressed in terms of vector spherical harmonics (see Appendix \ref{app:definitions}). Using the property $ \hat {\bf G}_0({\vec r, \vec r_j+\vec r', \omega}) = \hat {\bf G}_0({\vec r-\vec r_j, \vec r', \omega}) $,  and substituting \eqref{ein} into \eqref{out}, we obtain

%\begin{widetext}
  \begin{alignat}{2}
\label{out80}
\vec E (\vec r)=E_0\sum_j\frac{i k_1^3  (\varepsilon_2-\varepsilon_1)}{(4\pi)^2}  \sum_{p_i, p_r, n, m}\sum_{p_i', p_r',n', m'}(2-\delta_0) \frac {2n+1}{n(n+1)} \frac {(n-m)!}{(n+m)!}  D_{ p_i' p_r' n' m'} \vec W^{(3)}_{p_ip_rmn} (k_1, \vec r-\vec r_j) \cdot  \nonumber \\
\cdot e^{i(\vec k_b \cdot \vec r_j)} \int\limits_{V}  d^3 \vec r' [ \vec W^{(1)}_{p_ip_rmn} (k_1, \vec r')\cdot \vec W^{(1)}_{p_i'p_r'm'n'} (k_2,\vec r') ]
\end{alignat}
%\end{widetext}
here  superscript $(3)$ stands for the outgoing spherical Hankel wave. Now we dwell on the case when the array is {composed of} spherical nanoparticles. Exploiting the VSH orthogonality properties, the integral over sphere can be taken analytically \cite{Stratton}. It is proportional to Kroneker delta $\delta_{p_ip'_i} \delta_{p_rp'_r}\delta_{mm'}\delta_{nn'}$ which removes one summation. Combining all coefficients including the intergral into $\tilde D$ we obtain that the field outside the array is expressed with the formula:

  \begin{alignat}{1}
\label{out000}
\vec E (\vec r)=E_0\sum_{^{  p_i, p_r,m,n}_{ \ \  \ \ j}}  \tilde D_{p_i p_rmn}\vec W^{(3)}_{p_ip_rmn} (k_1, \vec r-\vec r_j) \cdot
 e^{i(\vec k_b \cdot \vec r_j)}.
\end{alignat}
Note, that coefficient $\tilde D_{p_ip_rmn}$ is non-zero only if the harmonic with such numbers is presented in the field expansion inside the sphere. If the fields created by each unit cell are already known, we can also start the considerations from this formula.

%To find the sum over $j$ we use the Weyl identity for  VSH expansion through the plane waves \cite{wittmann,stout} (see Appendix \ref{Weyl}) and obtain the formula:

To obtain formula \eqref{eq:out202} we exploit the Weyl identity for  VSH expansion through the plane waves in case when $z_n(kr)$ is replaced by spherical hankel functions\cite{wittmann,stout}:
\begin{equation}
\vec M^{(3)}_{p_rmn}(k, \vec r) =\frac{i^{-n}}{2\pi k} \iint  \limits_{-\infty}^{\infty} dk_{||}\frac{e^{ i{(k_xx+k_yy\pm k_zz)}}}{k_z}\left[\vec X_{p_rmn}\left(\frac{\vec k}{k}\right)\right] ,%e^{i\vec k \vec r}
\end{equation}

\begin{equation} \label{plane}
\vec N^{(3)}_{p_rmn}(k, \vec r) =\frac{i^{-n}}{2\pi k}  \iint  \limits_{-\infty}^{\infty} dk_{||} \frac{e^{ i{(k_xx+k_yy\pm k_zz)}}}{k_z}\left[\vec Z_{p_rmn}\left(\frac{\vec k}{k}\right)\right] .%e^{i\vec k \vec r}
\end{equation}

Sign before $k_z$ is depends on z- sign.
Redefining the harmonics, we have
\begin{equation} \label{plane}
\vec W^{(3)}_{p_ip_rmn}(k, \vec r) =\frac{i^{-n}}{2\pi k}  \iint  \limits_{-\infty}^{\infty} dk_{||} \frac{e^{ i{(k_xx+k_yy\pm k_zz)}}}{k_z}\left[\vec Y_{p_ip_rmn}\left(\frac{\vec k}{k}\right)\right], %e^{i\vec k \vec r}
\end{equation}
where $k_z=\sqrt{k^2-k_x^2-k_y^2}$. Substituting this expansion into \eqref{out000}, we get

 \begin{alignat}{2}
\label{out20}
\vec E (\vec r)=E_0\sum_{j, p_i, p_r, n, m}  \tilde D_{ p_i p_rmn}\cdot
 e^{i(\vec k_b \cdot \vec r_j)}
 \cdot \frac{i^{-n}}{2\pi k_1} \iint  \limits_{-\infty}^{\infty} dk_x dk_y \frac{e^{ i{(\vec k (\vec r - \vec r_j))}}}{k_z}\left[\vec Y_{p_ip_rmn}\left(\frac{\vec k}{k}\right)\right].
\end{alignat}
This expansion helps us to apply the summation formula
\begin{equation} \label{summ}
\sum_{j} e^{i \vec k \vec x_j}=V_{b}\sum_{\vec K} \delta({\vec k-\vec K})
\end{equation}
where $\vec K$ is reciprocal lattice vector and $V_b$ is the volume of the Brillouin zone  ($\frac{(2\pi)^2}{d^2}$ for square lattice), and substituting \eqref{summ} into \eqref{out20}, we obtain \eqref{eq:out202}.

 % \begin{alignat}{3}
%\vec E (\vec r)=E_0 \frac{2 \pi }{{d^2 k_1 k_z}}
 %{e^{ i{\vec k_1 \vec r}}}
% \cdot \sum_{p_r,m,n}  \tilde D^X_{p_rmn}\left[\vec X_{p_rmn}\left(\frac{\vec k_1}{k_1}\right)\right] +i \tilde D^Z_{p_rmn}\left[\vec Z_{p_rmn}\left(\frac{\vec k_1}{k_1}\right)\right]
% \end{alignat}
%now all coefficients $\tilde D^Y_{p_rmn} $ are real, which means that all multipolar contributions are in phase or antiphase (we recall that $\vec Z$ are pure imaginary (see Appendix \ref{app:definitions})).

\section{Relation between the coefficients $D$ and $\tilde D$} \label{appcoef}

The coefficients $D$ in \eqref{out80} and $\tilde D$ \eqref{out000} are connected by the formula
  \begin{alignat}{2}
\tilde D_{p_ip_rmn}=\frac{i k_1^3}{(4\pi)^2}(2-\delta_0) \frac {2n+1}{n(n+1)} \frac {(n-m)!}{(n+m)!}  D_{p_ip_rmn} \sum_{p_i', p_r',n', m'}\int\limits_{V}  d^3 \vec r'  (\varepsilon_2-\varepsilon_1) [ \vec W^{(1)}_{p_ip_rmn} (k_1, \vec r')\cdot \vec W^{(1)}_{p_i'p_r'm'n'} (k_2,\vec r') ] %\int\limits_{0}^a  d^3  r'^2 r' \Delta\varepsilon [ F(r') ]
\end{alignat}

% \begin{figure}[h!]
%	\includegraphics[width=0.25\linewidth]{strangefig-eps-converted-to.pdf}
%	 \caption{Illistration of the sphere circumscribed of a unit cell }
%\end{figure}

This formula describes both array of nanoparticles, and any photonic crystal slab, but in case of the array of spheres, the integral can be easily taken analytically.

 {\bf Remark} {\it Here we give the dyadic Green's function $ {\vec{ \hat{G}_0}}(\vec r, \vec r')$ only for the case when $r>r'$. In case of photonic crystal slab or non-spherical particles, we also need the part of Green's function when $r'<r$ \cite{Yee1994} to obtain the near-field. This will alter the intermediate calculations, since we have to compute the lattice sum for VSHs with spherical bessel functions. Nevertheless, the answer will have the same form.}

 We apply the orthogonality properties of vector spherical harmonics \cite{Stratton} [p.418], and consider integrals of magnetic and electric harmonics separately. %with help of the formula from wolfram http://functions.wolfram.com/Bessel-TypeFunctions/SphericalBesselJ/21/01/04/01/02/01/0002/ in case of spherical nanoparticles.
Implementing the angular integration, we reduce the integral to the integral of r-dependent Bessel functions products, which also can be computed analytically.
For magnetic harmonics we have
  \begin{alignat}{2}
\tilde D^M_{p_ip_rmn}=iD^M_{p_ip_rmn}a^2\varepsilon_1 [k_1^2 j_{n-1}(k_1a) j_{n}(k_2a)-k_1k_2 j_{n-1}(k_2a) j_{n}(k_1a)]%\int\limits_{0}^a  d^3  r'^2 r' \Delta\varepsilon [ F(r') ]
\end{alignat}
where $a$ is nanoparticle radius, and for electric
  \begin{alignat}{2}
\tilde D^N_{p_ip_rmn}=iD^N_{p_ip_rmn}a^2\varepsilon_1\left(\frac{n+1}{2n+1} [k_1^2 j_{n-2}(k_1a) j_{n-1}(k_2a)-k_1k_2 j_{n-2}(k_2a) j_{n-1}(k_1a)]+\right. \nonumber \\
+\left. \frac{n}{2n+1} [k_1^2 j_{n}(k_1a) j_{n+1}(k_2a)-k_1k_2 j_{n}(k_2a) j_{n+1}(k_1a)]\right)%\int\limits_{0}^a  d^3  r'^2 r' \Delta\varepsilon [ F(r') ]
\end{alignat}

Note that this expression turns to zero at some frequencies, so we can have zero $\tilde D$ when $D$ is non-zero. This refers to the anapoles of the spherical nanoparticles. The frequency where the anapole appears is the same as for single isolated nanoparticle.

If we have other type of surface, for example, photonic crystal slab with holes, or array of the cylinders, the orthogonality property can't be applied and the integral $\int\limits_{V}  d^3 \vec r' \Delta\varepsilon [ \vec W^{(1)}_{p_ip_rmn} (k_1, \vec r')\cdot \vec W^{(1)}_{p_i'p_r'm'n'} (k_2,\vec r') ] $ will mix some harmonics. However, all the harmonics, which admix, are already  presented in the expansion of the field inside the cell. This will just alter the coefficients before the outgoing multipoles, but not the  multipolar content.

%

 %are the scalar spherical functions, proportional to the tesseral spherical functions. Functions $z_n(\rho)$ can be replaced by  spherical Bessel functions of any type.

%\begin{gather}
%{\bf \vec{M}}_{emn}(k, \vec r) = \frac {-m}{\sin(\theta)}\sin(m\phi)P_n^m(\cos(\theta))z_n(\rho){\bf \vec e_{\theta}}- \nonumber \\
%-\cos(m\phi) \frac{dP_n^m(\cos(\theta))}{d\theta}z_n(\rho){\bf \vec e_{\phi}}\:,
%\end{gather}
%
%\begin{gather}
%{\bf \vec{M}}_{omn}(k, \vec r) = \frac {m}{\sin(\theta)}\cos(m\phi)P_n^m(\cos(\theta))z_n(\rho){\bf\vec e_{\theta}}- \nonumber\\
%-\sin(m\phi) \frac{dP_n^m(\cos(\theta))}{d\theta}z_n(\rho){\bf\vec e_{\phi}}\:,
%\end{gather}
%
%\begin{gather}
%{\bf \vec{N}}_{emn}(k, \vec r) = \frac {z_n(\rho)}{\rho}\cos(m\phi)n(n+1)P_n^m(\cos(\theta)){\bf\vec e_{r}}+ \nonumber\\
%+\cos(m\phi) \frac{dP_n^m(\cos(\theta))}{d\theta}\frac1{\rho}\frac{d}{d\rho}[\rho z_n(\rho)]{\bf\vec e_{\theta}}-\\-
%m\sin(m\phi) \frac{P_n^m(\cos(\theta))}{\sin(\theta)}\frac1{\rho}\frac{d}{d\rho}[\rho z_n(\rho)]{\bf\vec e_{\phi}} \:,\nonumber
%\end{gather}
%
%\begin{gather}
%{\bf \vec{N}}_{omn}(k, \vec r) = \frac {z_n(\rho)}{\rho}\sin(m\phi)n(n+1)P_n^m(\cos(\theta)){\bf\vec e_{r}}+ \nonumber \\+
%\sin(m\phi) \frac{dP_n^m(\cos(\theta))}{d\theta}\frac1{\rho}\frac{d}{d\rho}[\rho z_n(\rho)]{\bf\vec e_{\theta}}+\\+
%m\cos(m\phi) \frac{P_n^m(\cos(\theta))}{\sin(\theta)}\frac1{\rho}\frac{d}{d\rho}[\rho z_n(\rho)]{\bf\vec e_{\phi}}\:.\nonumber
%\end{gather}
 %\newpage

 \end{widetext}

\end{document}